\documentclass[12pt,oneside,reqno,american]{amsart}
\usepackage{biolinum}
\usepackage[T1]{fontenc}
\usepackage[latin9]{inputenc}
\usepackage{color}
\definecolor{lyxboxbgcolor}{rgb}{1, 0, 0}
\definecolor{lyxnotefontcolor}{rgb}{0.75, 0.75, 0.75}
\usepackage{verbatim}
\usepackage{booktabs}
\usepackage{amsbsy}
\usepackage{amstext}
\usepackage{amsthm}
\usepackage{amssymb}
\usepackage{xargs}[2008/03/08]

\makeatletter

\providecommand{\tabularnewline}{\\}

\numberwithin{equation}{section}
\numberwithin{figure}{section}
\numberwithin{table}{section}

\usepackage{mathtools}
\usepackage{amsthm}
\usepackage{pdfsync} 
\usepackage{hyperref}
\usepackage[all]{xy}

\usepackage[T1]{fontenc}
\usepackage{Alegreya} 
\usepackage[stix2,scaled=0.96]{newtxmath}
\usepackage[cal=boondoxo,bb=boondox,frak=boondox,scaled=1.10]{mathalfa}
\usepackage[scaled=.9]{rsfso}
\usepackage{xcolor} 
\definecolor{brown(traditional)}{rgb}{0.59, 0.29, 0.0}
\definecolor{blue(ryb)}{rgb}{0.01, 0.28, 1.0}
\definecolor{red}{rgb}{1.0, 0.0, 0.0}
\definecolor{magenta}{rgb}{1.0, 0.0, 1.0}
\definecolor{mahogany}{rgb}{0.75, 0.25, 0.0}
\definecolor{lavenderpurple}{rgb}{0.59, 0.48, 0.71}
\definecolor{olive}{rgb}{0.5, 0.5, 0.0}
\definecolor{brickred}{rgb}{0.8, 0.25, 0.33}
\definecolor{antiquefuchsia}{rgb}{0.57, 0.36, 0.51}
\definecolor{bole}{rgb}{0.47, 0.27, 0.23}
\definecolor{darkolivegreen}{rgb}{0.33, 0.42, 0.18}
\definecolor{deepjunglegreen}{rgb}{0.0, 0.29, 0.29}
\definecolor{brickred}{rgb}{0.8, 0.25, 0.33}
\definecolor{deepjunglegreen}{rgb}{0.0, 0.29, 0.29}
\definecolor{darkpastelgreen}{rgb}{0.01, 0.75, 0.24}
\definecolor{green(pigment)}{rgb}{0.0, 0.65, 0.31}
\definecolor{junglegreen}{rgb}{0.16, 0.67, 0.53}
\definecolor{officegreen}{rgb}{0.0, 0.5, 0.0}
\definecolor{seagreen}{rgb}{0.18, 0.55, 0.34}
\definecolor{teal}{rgb}{0.0, 0.5, 0.5}
\definecolor{brightgreen}{rgb}{0.4, 1.0, 0.0}
\definecolor{electricgreen}{rgb}{0.0, 1.0, 0.0}
\definecolor{malachite}{rgb}{0.04, 0.85, 0.32}

\usepackage{accents}
\usepackage{colortbl}
\usepackage{verbatim}
\usepackage{float}
\usepackage{booktabs}
\usepackage{tikz}
\usetikzlibrary{arrows.meta,positioning}
\tikzset{
  negchg/.style={circle,draw,fill=cyan!8,inner sep=2.2pt},
  poschg/.style={circle,draw,fill=red!25,inner sep=2.2pt},
  vec/.style={-{Stealth[length=2.2mm]},thick},
  frc/.style={-{Stealth[length=2.2mm]},thick,red!70!black},
  vel/.style={-{Stealth[length=2.2mm]},thick,blue!80!black}
}

\makeatother

\providecommand\theoremname{Theorem}
\theoremstyle{plain}
\newtheorem{thm}{\protect\theoremname}[section]
\providecommand\lemmaname{Lemma}
\newtheorem{lem}[thm]{\protect\lemmaname}
\providecommand\assumptionname{Assumption}
\newtheorem{assumption}[thm]{\protect\assumptionname}
\providecommand\remarkname{Remark}
\theoremstyle{remark}
\newtheorem{rem}[thm]{\protect\remarkname}
\providecommand\propositionname{Proposition}
\theoremstyle{plain}
\newtheorem{prop}[thm]{\protect\propositionname}
\providecommand\corollaryname{Corollary}
\newtheorem{cor}[thm]{\protect\corollaryname}
\providecommand\definitionname{Definition}
\theoremstyle{definition}
\newtheorem{defn}[thm]{\protect\definitionname}
\usepackage{babel}
\begin{document}

\global\long\def\ga{\alpha}%
\global\long\def\gb{\beta}%
\global\long\def\ggm{\gamma}%
\global\long\def\go{\omega}%
\global\long\def\gs{\sigma}%
\global\long\def\gd{\delta}%
\global\long\def\gD{\Delta}%
\global\long\def\vph{\phi}%
\global\long\def\gf{\phi}%
\global\long\def\gk{\kappa}%
\global\long\def\gl{\lambda}%
\global\long\def\gz{\zeta}%
\global\long\def\gh{\eta}%
\global\long\def\gy{\upsilon}%
\global\long\def\gth{\theta}%
\global\long\def\gO{\Omega}%
\global\long\def\gG{\Gamma}%

\global\long\def\eps{\varepsilon}%
\global\long\def\epss#1#2{\varepsilon^{#1}_{#2}}%
\global\long\def\ep#1{\eps_{#1}}%

\global\long\def\wh#1{\widehat{#1}}%
\global\long\def\hi{\hat{\imath}}%
\global\long\def\hj{\hat{\jmath}}%
\global\long\def\hk{\hat{k}}%
\global\long\def\ol#1{\overline{#1}}%
\global\long\def\ul#1{\underline{#1}}%

\global\long\def\spec#1{\textsf{#1}}%

\global\long\def\v#1{\boldsymbol{#1}}%

\global\long\def\ui{\wh{\boldsymbol{\imath}}}%
\global\long\def\uj{\wh{\boldsymbol{\jmath}}}%
\global\long\def\uk{\widehat{\boldsymbol{k}}}%

\global\long\def\uI{\widehat{\mathbf{I}}}%
\global\long\def\uJ{\widehat{\mathbf{J}}}%
\global\long\def\uK{\widehat{\mathbf{K}}}%

\global\long\def\mc#1{\mathcal{#1}}%
\global\long\def\bs#1{\boldsymbol{#1}}%
\global\long\def\vect#1{\mathbf{#1}}%
\global\long\def\bi#1{\textbf{\emph{#1}}}%

\global\long\def\uv#1{\widehat{\boldsymbol{#1}}}%
\global\long\def\cross{\times}%

\global\long\def\di{d}%
\global\long\def\dee#1{\mathop{d#1}}%

\global\long\def\ddt{\frac{\dee{}}{\dee t}}%
\global\long\def\dbyd#1{\frac{\dee{}}{\dee{#1}}}%
\global\long\def\dby#1#2{\frac{\partial#1}{\partial#2}}%
\global\long\def\dxdt#1{\frac{\dee{#1}}{\dee t}}%

\global\long\def\vct#1{\bs{#1}}%

\global\long\def\partialby#1#2{\frac{\partial#1}{\partial x^{#2}}}%
\newcommandx\parder[2][usedefault, addprefix=\global, 1=]{\frac{\partial#2}{\partial#1}}%
\global\long\def\supdot{^{\!\bs{\mathord{\cdot}}}}%

\global\long\def\fall{,\quad\text{for all}\quad}%

\global\long\def\reals{\mathbb{R}}%

\global\long\def\rthree{\reals^{3}}%
\global\long\def\rsix{\reals^{6}}%
\global\long\def\rn{\reals^{n}}%
\global\long\def\eucl{\mathbb{E}}%
\global\long\def\euthree{\eucl^{3}}%
\global\long\def\euln{\eucl^{n}}%

\global\long\def\prn{\reals^{n+}}%
\global\long\def\nrn{\reals^{n-}}%
\global\long\def\cprn{\overline{\reals}^{n+}}%
\global\long\def\cnrn{\overline{\reals}^{n-}}%
\global\long\def\rt#1{\reals^{#1}}%
\global\long\def\rtw{\reals^{12}}%

\global\long\def\les{\leqslant}%
\global\long\def\ges{\geqslant}%

\global\long\def\dX{\dee{\bp}}%
\global\long\def\dx{\dee x}%
\global\long\def\D{D}%

\global\long\def\from{\colon}%
\global\long\def\tto{\longrightarrow}%
\global\long\def\lmt{\longmapsto}%
\global\long\def\lhr{\lhook\joinrel\longrightarrow}%
\global\long\def\mto{\mapsto}%

\global\long\def\abs#1{\left|#1\right|}%

\global\long\def\isom{\cong}%

\global\long\def\comp{\circ}%

\global\long\def\cl#1{\overline{#1}}%

\global\long\def\fun{\varphi}%

\global\long\def\interior{\textrm{Int}\,}%
\global\long\def\inter#1{\kern0pt  #1^{\mathrm{o}}}%
\global\long\def\interior{\textrm{Int}\,}%
\global\long\def\inter#1{\kern0pt  #1^{\mathrm{o}}}%
\global\long\def\into{\mathrm{o}}%

\global\long\def\sign{\textrm{sign}\,}%
\global\long\def\sgn#1{(-1)^{#1}}%
\global\long\def\sgnp#1{(-1)^{\abs{#1}}}%

\global\long\def\du#1{#1^{*}}%

\global\long\def\tsum{{\textstyle \sum}}%
\global\long\def\lsum{{\textstyle \sum}}%

\global\long\def\dimension{\textrm{dim}\,}%

\global\long\def\esssup{\textrm{ess}\,\sup}%

\global\long\def\ess{\textrm{{ess}}}%

\global\long\def\kernel{\mathop{\textrm{\textup{Kernel}}}}%

\global\long\def\support{\mathop{\textrm{\textup{supp}}}}%

\global\long\def\image{\mathop{\textrm{\textup{Image}}}}%

\global\long\def\diver{\mathop{\textrm{\textup{div}}}}%

\global\long\def\spanv{\textrm{span}}%

\global\long\def\tr{\mathop{\textrm{\textup{tr}}}}%
\global\long\def\tran{\mathrm{tr}}%

\global\long\def\opt{\mathrm{opt}}%

\global\long\def\incl{\mathcal{I}}%
\global\long\def\iden{\imath}%
\global\long\def\idnt{\textrm{Id}}%
\global\long\def\rest{\rho}%
\global\long\def\extnd{e_{0}}%

\global\long\def\proj{\textrm{pr}}%

\global\long\def\L#1{L\bigl(#1\bigr)}%
\global\long\def\LS#1{L_{S}\bigl(#1\bigr)}%

\global\long\def\ino#1{\int_{#1}}%

\global\long\def\half{\frac{1}{2}}%
\global\long\def\shalf{{\scriptstyle \half}}%
\global\long\def\third{\frac{1}{3}}%

\global\long\def\empt{\varnothing}%

\global\long\def\innp#1#2{\left\langle #1,#2\right\rangle }%

\global\long\def\resto#1{|_{#1}}%
\global\long\def\compat#1#2{\left.#1\right|_{#2}}%

\global\long\def\paren#1{\left(#1\right)}%
\global\long\def\bigp#1{\bigl(#1\bigr)}%
\global\long\def\biggp#1{\biggl(#1\biggr)}%
\global\long\def\Bigp#1{\Bigl(#1\Bigr)}%

\global\long\def\braces#1{\left\{  #1\right\}  }%
\global\long\def\sqbr#1{\left[#1\right]}%
\global\long\def\anglep#1{\left\langle #1\right\rangle }%

\global\long\def\bigabs#1{\bigl|#1\bigr|}%
\global\long\def\dotp#1{#1^{\centerdot}}%
\global\long\def\pdot#1{#1^{\bs{\!\cdot}}}%

\global\long\def\eq{\sim}%
\global\long\def\quot{/\!\!\eq}%
\global\long\def\by{\!/\!}%

\global\long\def\stp{\text{{\small \ensuremath{\bigodot}}}}%
\global\long\def\tp{\text{{\small \ensuremath{\bigotimes}}}}%

\global\long\def\mi#1{#1}%
\global\long\def\mii{I}%
\global\long\def\mie#1#2{#1_{1}\cdots#1_{#2}}%

\global\long\def\smi#1{\boldsymbol{#1}}%
\global\long\def\asmi#1{#1}%
\global\long\def\ordr#1{\left\langle #1\right\rangle }%

\global\long\def\symm#1{\paren{#1}}%
\global\long\def\smtr{\mathcal{S}}%

\global\long\def\perm{p}%
\global\long\def\sperm{\mathcal{P}}%

\global\long\def\oneto{1,\dots,}%

\global\long\def\lisub#1#2#3{#1_{1}#2\dots#2#1_{#3}}%

\global\long\def\lisup#1#2#3{#1^{1}#2\dots#2#1^{#3}}%

\global\long\def\lisubb#1#2#3#4{#1_{#2}#3\dots#3#1_{#4}}%

\global\long\def\lisubbc#1#2#3#4{#1_{#2}#3\cdots#3#1_{#4}}%

\global\long\def\lisubbwout#1#2#3#4#5{#1_{#2}#3\dots#3\widehat{#1}_{#5}#3\dots#3#1_{#4}}%

\global\long\def\lisubc#1#2#3{#1_{1}#2\cdots#2#1_{#3}}%

\global\long\def\lisupc#1#2#3{#1^{1}#2\cdots#2#1^{#3}}%

\global\long\def\lisupp#1#2#3#4{#1^{#2}#3\dots#3#1^{#4}}%

\global\long\def\lisuppc#1#2#3#4{#1^{#2}#3\cdots#3#1^{#4}}%

\global\long\def\lisuppwout#1#2#3#4#5#6{#1^{#2}#3#4#3\wh{#1^{#6}}#3#4#3#1^{#5}}%

\global\long\def\lisubbwout#1#2#3#4#5#6{#1_{#2}#3#4#3\wh{#1}_{#6}#3#4#3#1_{#5}}%

\global\long\def\lisubwout#1#2#3#4{#1_{1}#2\dots#2\widehat{#1}_{#4}#2\dots#2#1_{#3}}%

\global\long\def\lisupwout#1#2#3#4{#1^{1}#2\dots#2\widehat{#1^{#4}}#2\dots#2#1^{#3}}%

\global\long\def\lisubwoutc#1#2#3#4{#1_{1}#2\cdots#2\widehat{#1}_{#4}#2\cdots#2#1_{#3}}%

\global\long\def\twp#1#2#3{\dee{#1}^{#2}\wedge\dee{#1}^{#3}}%

\global\long\def\thp#1#2#3#4{\dee{#1}^{#2}\wedge\dee{#1}^{#3}\wedge\dee{#1}^{#4}}%

\global\long\def\fop#1#2#3#4#5{\dee{#1}^{#2}\wedge\dee{#1}^{#3}\wedge\dee{#1}^{#4}\wedge\dee{#1}^{#5}}%

\global\long\def\idots#1{#1\dots#1}%
\global\long\def\icdots#1{#1\cdots#1}%

\global\long\def\norm#1{\|#1\|}%

\global\long\def\nonh{\heartsuit}%

\global\long\def\nhn#1{\norm{#1}^{\nonh}}%

\global\long\def\bigmid{\,\bigl|\,}%

\global\long\def\trps{^{{\scriptscriptstyle \textsf{T}}}}%

\global\long\def\testfuns{\mathcal{D}}%

\global\long\def\ntil#1{\tilde{#1}{}}%

\global\long\def\pis{y}%
\global\long\def\xo{\pis_{0}}%
\global\long\def\x{x}%

\global\long\def\pib{x}%
\global\long\def\bp{X}%
\global\long\def\ii{i}%
\global\long\def\ia{\alpha}%
\global\long\def\fp{y}%
\global\long\def\piv{v}%

\global\long\def\ib{i}%
\global\long\def\is{\alpha}%

\global\long\def\pbndo{\Gamma}%
\global\long\def\bndoo{\pbndo_{0}}%
 
\global\long\def\bndot{\pbndo_{t}}%
\global\long\def\intb{\inter{\body}}%
\global\long\def\bndb{\bdry\body}%

\global\long\def\cloo{\cl{\gO}}%

\global\long\def\nor{\nu}%
\global\long\def\Nor{\mathbf{N}}%

\global\long\def\dA{\dee A}%

\global\long\def\dV{\dee V}%

\global\long\def\eps{\varepsilon}%

\global\long\def\tv{v}%
\global\long\def\av{u}%

\global\long\def\svs{\mathcal{W}}%
\global\long\def\vs{\mathbf{V}}%
\global\long\def\avs{\mathbf{U}}%
\global\long\def\affsp{\mathcal{A}}%
\global\long\def\man{\mathcal{M}}%
\global\long\def\odman{\mathcal{N}}%
\global\long\def\subman{\mathcal{V}}%
\global\long\def\pt{p}%

\global\long\def\vbase{e}%
\global\long\def\sbase{\v e}%
\global\long\def\msbase{\mathfrak{e}}%
\global\long\def\vect{v}%
\global\long\def\dbase{\sbase}%

\global\long\def\chart{\varphi}%
\global\long\def\Chart{\Phi}%

\global\long\def\mind{\alpha}%
\global\long\def\vb{W}%
\global\long\def\vbp{\pi}%

\global\long\def\vbt{\mathcal{E}}%
\global\long\def\fib{\vs}%
\global\long\def\vbts{W}%
\global\long\def\avb{U}%
\global\long\def\vbp{\xi}%

\global\long\def\chart{\vph}%
\global\long\def\vbchart{\Phi}%

\global\long\def\jetb#1{J^{#1}}%
\global\long\def\jet#1{j^{1}(#1)}%
\global\long\def\tjet{\tilde{\jmath}}%

\global\long\def\Jet#1{J^{1}(#1)}%

\global\long\def\jetm{j}%

\global\long\def\coj{\mathfrak{d}}%

\global\long\def\alt{\mathfrak{A}}%

\global\long\def\pou{\eta}%

\global\long\def\ext{{\textstyle \bigwedge}}%
\global\long\def\forms{\Omega}%

\global\long\def\dotwedge{\dot{\mbox{\ensuremath{\wedge}}}}%

\global\long\def\vel{\theta}%

\global\long\def\Jac{\mathcal{J}}%

\global\long\def\contr{\mathbin{\raisebox{0.4pt}{\mbox{\ensuremath{\lrcorner}}}}}%
\global\long\def\fcor{\llcorner}%
\global\long\def\bcor{\lrcorner}%
\global\long\def\fcontr{\mathbin{\raisebox{0.4pt}{\mbox{\ensuremath{\llcorner}}}}}%

\global\long\def\lie{\mathcal{L}}%

\global\long\def\ssym#1#2{\ext^{#1}T^{*}#2}%

\global\long\def\sh{^{\sharp}}%

\global\long\def\nfo{\ext^{n}T^{*}\base}%
\global\long\def\dfs{\ext^{d}T^{*}\base}%
\global\long\def\dmfs{\ext^{d-1}T^{*}\base}%

\global\long\def\spc{\mathcal{S}}%
\global\long\def\sptm{\mathcal{E}}%
\global\long\def\evnt{e}%
\global\long\def\frame{\Psi}%

\global\long\def\timeman{\mathcal{T}}%
\global\long\def\zman{t}%
\global\long\def\dims{n}%
\global\long\def\m{\dims-1}%
\global\long\def\dimw{m}%

\global\long\def\wc{z}%

\global\long\def\fourv#1{\mbox{\ensuremath{\mathfrak{#1}}}}%

\global\long\def\body{\mathcal{B}}%
\global\long\def\man{\mathcal{M}}%
\global\long\def\var{\mathcal{V}}%
\global\long\def\base{\mathcal{X}}%
\global\long\def\fb{\mathcal{Y}}%
\global\long\def\srfc{\mathcal{Z}}%
\global\long\def\dimb{n}%
\global\long\def\dimf{m}%
\global\long\def\afb{\mathcal{Z}}%

\global\long\def\bdry{\partial}%

\global\long\def\gO{\varOmega}%

\global\long\def\reg{\gO}%
\global\long\def\bdrr{\bdry\reg}%

\global\long\def\bdom{\bdry\gO}%

\global\long\def\bndo{\partial\gO}%

\global\long\def\tpr{\vartheta}%

\global\long\def\mot{M}%
\global\long\def\vf{w}%
\global\long\def\const{h}%

\global\long\def\avf{u}%

\global\long\def\stn{\varepsilon}%
\global\long\def\djet{\chi}%

\global\long\def\jvf{\eps}%

\global\long\def\rig{r}%

\global\long\def\rigs{\mathcal{R}}%

\global\long\def\qrigs{\!/\!\rigs}%

\global\long\def\qd{\!/\,\!\kernel\diffop}%

\global\long\def\dis{\chi}%
\global\long\def\conf{\kappa}%
\global\long\def\invc{\hat{\conf}^{-1}}%
\global\long\def\dinvc{\hat{\conf}^{-1*}}%
\global\long\def\csp{\mathcal{Q}}%

\global\long\def\embds{\textrm{Emb}}%

\global\long\def\lc{A}%

\global\long\def\lv{\dot{A}}%
\global\long\def\alv{\dot{B}}%

\global\long\def\j{\mathop{\mathrm{j}}}%
\global\long\def\mapp{M}%
\global\long\def\J{J}%
\global\long\def\jex{\mathop{}\!\mathrm{j}}%

\global\long\def\fc{F}%
\global\long\def\load{f}%
\global\long\def\afc{g}%

\global\long\def\bfc{\mathbf{b}}%
\global\long\def\bfcc{b}%

\global\long\def\sfc{\mathbf{t}}%
\global\long\def\sfcc{t}%

\global\long\def\stm{\varsigma}%
\global\long\def\std{S}%
\global\long\def\tst{\sigma}%
\global\long\def\tstd{s}%
\global\long\def\st{\sigma}%
\global\long\def\vst{\varsigma}%
\global\long\def\vstd{S}%
\global\long\def\tstm{\sigma}%
\global\long\def\vstm{\varsigma}%

\global\long\def\stp{S_{P}}%
\global\long\def\slf{R}%

\global\long\def\crel{\Phi}%

\global\long\def\stmat{\tau}%

\global\long\def\gdiv{\bdry\textrm{iv\,}}%
\global\long\def\extjet{\mathfrak{d}}%

\global\long\def\smc#1{\mathfrak{#1}}%

\global\long\def\nhs{P}%
\global\long\def\nhsa{P}%
\global\long\def\nhsb{\underline{P}}%

\global\long\def\soc{Z}%

\global\long\def\sts{\varSigma}%
\global\long\def\spstd{\mathfrak{S}}%
\global\long\def\sptst{\mathfrak{T}}%
\global\long\def\spnhs{\mathcal{P}}%
\global\long\def\Ljj{\L{J^{1}(J^{k-1}\vb),\ext^{n}T^{*}\base}}%

\global\long\def\spsb{\text{{\Large \ensuremath{\Delta}}}}%

\global\long\def\ened{\mathfrak{w}}%
\global\long\def\energy{\mathfrak{W}}%

\global\long\def\ebdfc{T}%
\global\long\def\optimum{\st^{\textrm{opt}}}%
\global\long\def\scf{K}%

\global\long\def\grp{G}%
\global\long\def\gact{A}%
\global\long\def\gid{e}%
\global\long\def\gel{\ggm}%

\global\long\def\ael{\upsilon}%
\global\long\def\lal{\mathfrak{g}}%

\global\long\def\expr{\Pi}%

\global\long\def\aprop{Q}%

\global\long\def\flux{\omega}%
\global\long\def\aflux{\psi}%

\global\long\def\fform{\tau}%

\global\long\def\dimn{n}%

\global\long\def\sdim{{\dimn-1}}%

\global\long\def\fdens{\phi}%

\global\long\def\pform{s}%
\global\long\def\vform{\beta}%
\global\long\def\sform{\tau}%
\global\long\def\flow{\vf}%
\global\long\def\n{\m}%
\global\long\def\cmap{\mathfrak{t}}%
\global\long\def\vcmap{\varSigma}%

\global\long\def\mvec{\mathfrak{v}}%
\global\long\def\mveco#1{\mathfrak{#1}}%
\global\long\def\mv#1{\mathfrak{#1}}%
\global\long\def\smbase{\mathfrak{e}}%
\global\long\def\spx{\simp}%
\global\long\def\il{l}%
\global\long\def\awe{\frown}%

\global\long\def\hp{H}%
\global\long\def\ohp{h}%

\global\long\def\hps{G_{\dims-1}(T\spc)}%
\global\long\def\ohps{G^{\perp}_{\dims-1}(T\spc)}%

\global\long\def\hyper{\mathcal{S}}%

\global\long\def\hpsx{G_{\dims-1}(\tspc)}%
\global\long\def\ohpsx{G^{\perp}_{\dims-1}(\tspc)}%

\global\long\def\fbun{F}%

\global\long\def\flowm{\Phi}%

\global\long\def\tgb{T\spc}%
\global\long\def\ctgb{T^{*}\spc}%
\global\long\def\tspc{T_{\pis}\spc}%
\global\long\def\dspc{T^{*}_{\pis}\spc}%

\global\long\def\fflow{\fourv J}%
\global\long\def\fvform{\mathfrak{b}}%
\global\long\def\fsform{\mathfrak{t}}%
\global\long\def\fpform{\mathfrak{s}}%
\global\long\def\lfc{\mathfrak{F}}%

\global\long\def\maxw{\mathfrak{g}}%
\global\long\def\frdy{\mathfrak{f}}%
\global\long\def\ptnl{\gf}%
\global\long\def\pts{\Psi}%
\global\long\def\tptn{\Psi}%
\global\long\def\vptn{\mathfrak{a}}%
\global\long\def\mtst{\tstd_{M}}%
\global\long\def\mvst{\vstd_{M}}%

\global\long\def\sobp#1#2{W^{#1}_{#2}}%

\global\long\def\inner#1#2{\left\langle #1,#2\right\rangle }%

\global\long\def\fields{\sobp pk(\vb)}%

\global\long\def\bodyfields{\sobp p{k_{\partial}}(\vb)}%

\global\long\def\forces{\sobp pk(\vb)^{*}}%

\global\long\def\bfields{\sobp p{k_{\partial}}(\vb\resto{\bndo})}%

\global\long\def\loadp{(\sfc,\bfc)}%

\global\long\def\strains{\lp p(\jetb k(\vb))}%

\global\long\def\stresses{\lp{p'}(\jetb k(\vb)^{*})}%

\global\long\def\diffop{D}%

\global\long\def\strainm{E}%

\global\long\def\incomps{\vbts_{\yieldf}}%

\global\long\def\devs{L^{p'}(\eta^{*}_{1})}%

\global\long\def\incompsns{L^{p}(\eta_{1})}%

\global\long\def\testf{\mathcal{D}}%
\global\long\def\dists{\mathcal{D}'}%

\global\long\def\codiv{\boldsymbol{\partial}}%

\global\long\def\currof#1{\tilde{#1}}%

\global\long\def\chn{c}%
\global\long\def\chnsp{\mathbf{C}}%

\global\long\def\current{T}%
\global\long\def\curr{R}%

\global\long\def\curd{S}%
\global\long\def\curwd#1{\wh{#1}}%
\global\long\def\curnd#1{\wh{#1}}%

\global\long\def\contrf{{\scriptstyle \smallfrown}}%

\global\long\def\prodf{{\scriptstyle \smallsmile}}%

\global\long\def\form{\omega}%

\global\long\def\dens{\rho}%

\global\long\def\simp{s}%
\global\long\def\ssimp{\Delta}%
\global\long\def\cpx{K}%

\global\long\def\cell{C}%

\global\long\def\chain{B}%
\global\long\def\A{A}%
\global\long\def\B{B}%

\global\long\def\ach{A}%

\global\long\def\coch{X}%

\global\long\def\scale{s}%

\global\long\def\fnorm#1{\norm{#1}^{\flat}}%

\global\long\def\chains{\mathcal{A}}%

\global\long\def\ivs{\boldsymbol{U}}%

\global\long\def\mvs{\boldsymbol{V}}%

\global\long\def\cvs{\boldsymbol{W}}%

\global\long\def\ndual#1{#1'}%

\global\long\def\nd{'}%

\global\long\def\cee#1{C^{#1}}%

\global\long\def\lone{\{L^{1}\}}%

\global\long\def\linf{L^{\infty}}%

\global\long\def\lp#1{L^{#1}}%

\global\long\def\ofbdo{(\bndo)}%

\global\long\def\ofclo{(\cloo)}%

\global\long\def\vono{(\gO,\rthree)}%

\global\long\def\lomu{\{L^{1,\mu}\}}%
\global\long\def\limu{L^{\infty,\mu}}%
\global\long\def\limub{\limu(\body,\rthree)}%
\global\long\def\lomub{\lomu(\body,\rthree)}%

\global\long\def\vonbdo{(\bndo,\rthree)}%
\global\long\def\vonbdoo{(\bndoo,\rthree)}%
\global\long\def\vonbdot{(\bndot,\rthree)}%

\global\long\def\vonclo{(\cl{\gO},\rthree)}%

\global\long\def\strono{(\gO,\reals^{6})}%

\global\long\def\sob{\{W^{1}_{1}\}}%

\global\long\def\sobb{\sob(\gO,\rthree)}%

\global\long\def\lob{\lone(\gO,\rthree)}%

\global\long\def\lib{\linf(\gO,\reals^{12})}%

\global\long\def\ofO{(\gO)}%

\global\long\def\oneo{{1,\gO}}%
\global\long\def\onebdo{{1,\bndo}}%
\global\long\def\info{{\infty,\gO}}%

\global\long\def\infclo{{\infty,\cloo}}%

\global\long\def\infbdo{{\infty,\bndo}}%
\global\long\def\lobdry{\lone(\bdry\gO,\rthree)}%

\global\long\def\ld{LD}%

\global\long\def\ldo{\ld\ofO}%
\global\long\def\ldoo{\ldo_{0}}%

\global\long\def\trace{\gamma}%
\global\long\def\dtrace{\delta}%
\global\long\def\gtrace{\beta}%

\global\long\def\pr{\proj_{\rigs}}%

\global\long\def\pq{\proj}%

\global\long\def\qr{\,/\,\reals}%

\global\long\def\aro{S_{1}}%
\global\long\def\art{S_{2}}%

\global\long\def\mo{m_{1}}%
\global\long\def\mt{m_{2}}%

\global\long\def\ebdfc{T}%

\global\long\def\mini{\Omega}%
\global\long\def\optimum{s^{\mathrm{opt}}}%
\global\long\def\scf{K}%
\global\long\def\opsf{\st^{\mathrm{opt}}}%
\global\long\def\doptimum{s^{\opt,{\scriptscriptstyle D}}}%
\global\long\def\loptimum{s^{\opt,{\scriptscriptstyle \mathcal{M}}}}%

\global\long\def\fsubs{M}%

\global\long\def\yieldc{B}%

\global\long\def\yieldf{Y}%

\global\long\def\trpr{\pi_{P}}%

\global\long\def\devpr{\pi_{\devsp}}%

\global\long\def\prsp{P}%

\global\long\def\devsp{D}%

\global\long\def\ynorm#1{\|#1\|_{\yieldf}}%

\global\long\def\colls{\Psi}%

\global\long\def\aro{S_{1}}%
\global\long\def\art{S_{2}}%

\global\long\def\mo{m_{1}}%
\global\long\def\mt{m_{2}}%

\global\long\def\trps{^{\mathsf{T}}}%

\global\long\def\hb{^{\mathrm{hb}}}%

\global\long\def\yieldst{s_{Y}}%

\global\long\def\yieldc{B}%

\global\long\def\lcap{C}%

\global\long\def\yieldf{Y}%

\global\long\def\sphpr{\pi_{P}}%

\global\long\def\devpr{\pi_{\devsp}}%

\global\long\def\prsp{P}%

\global\long\def\devsp{D}%

\global\long\def\ynorm#1{\|#1\|_{\yieldf}}%

\global\long\def\colls{\Psi}%

\global\long\def\cone{Q}%
\global\long\def\fpr{\Pi}%
\global\long\def\fprd{\fpr_{\devsp}}%
\global\long\def\fprp{\fpr_{\prsp}}%
\global\long\def\find{I_{\devsp}}%
\global\long\def\finp{I_{\prsp}}%
\global\long\def\fnorm#1{\norm{#1}_{\devsp}}%

\global\long\def\rig{r}%
\global\long\def\rigs{\mathcal{R}}%
\global\long\def\qrigs{\!/\!\rigs}%
\global\long\def\anv{\omega}%
\global\long\def\I{I}%
\global\long\def\mone{M_{1}}%

\global\long\def\bd{BD}%

\global\long\def\po{\proj_{0}}%
\global\long\def\normp#1{\norm{#1}'_{\ld}}%

\global\long\def\ssx{S}%

\global\long\def\smap{s}%

\global\long\def\smat{\chi}%

\global\long\def\sx{e}%

\global\long\def\snode{P}%
\global\long\def\newmacroname{\{\}}%

\global\long\def\elem{e}%

\global\long\def\nel{L}%

\global\long\def\el{l}%

\global\long\def\gr{g}%
\global\long\def\ngr{G}%

\global\long\def\eldof{\alpha}%

\global\long\def\glbs{\psi}%

\global\long\def\ipln{\phi}%

\global\long\def\ndof{D}%

\global\long\def\dof{d}%

\global\long\def\nldof{N}%

\global\long\def\ldof{n}%

\global\long\def\lvf{\chi}%

\global\long\def\amat{A}%
\global\long\def\bmat{B}%

\global\long\def\subsp{\mathcal{M}}%
\global\long\def\zerofn{Z}%

\global\long\def\snomat{E}%

\global\long\def\femat{E}%

\global\long\def\tmat{T}%

\global\long\def\fvec{f}%

\global\long\def\snsp{\mathcal{S}}%

\global\long\def\slnsp{\Phi}%
\global\long\def\dslnsp{\Phi^{{\scriptscriptstyle D}}}%

\global\long\def\ro{r_{1}}%

\global\long\def\rtwo{r_{2}}%

\global\long\def\rth{r_{3}}%

\global\long\def\fmax{M}%

\global\long\def\dform{\psi}%

\global\long\def\srfc{\mathcal{S}}%

\global\long\def\semib{\mathrm{SB}}%

\global\long\def\tm#1{\overrightarrow{#1}}%
\global\long\def\tmm#1{\underrightarrow{\overrightarrow{#1}}}%

\global\long\def\itm#1{\overleftarrow{#1}}%
\global\long\def\itmm#1{\underleftarrow{\overleftarrow{#1}}}%

\global\long\def\ptrac{\mathcal{P}}%

\global\long\def\nh#1{\hat{#1}}%
\global\long\def\nj{\hat{\jmath}}%
\global\long\def\nJ{\hat{J}}%
\global\long\def\rin#1{\mathfrak{#1}}%
\global\long\def\npi{\hat{\pi}}%
\global\long\def\rp{\rin p}%
\global\long\def\rq{\rin q}%
\global\long\def\rr{\rin r}%

\global\long\def\xty{(\base,\fb)}%
\global\long\def\xts{(\base,\spc)}%
\global\long\def\r{r}%
\global\long\def\ntm{(\reals^{n},\reals^{m})}%

\global\long\def\tproj{\frame_{\timeman}}%
\global\long\def\sproj{\frame_{\spc}}%

\global\long\def\cons{c}%
\global\long\def\optm{\go}%
\global\long\def\flxs{\mathcal{W}}%
\global\long\def\cost{Q}%

\global\long\def\mtn{e}%
\global\long\def\sppp{\lambda}%

\global\long\def\mtsp{\mathscr{E}}%

\global\long\def\disp{g}%
\global\long\def\diffs{G}%

\global\long\def\bv{BV}%

\global\long\def\Charge{Q}%

\global\long\def\pole{q}%

\global\long\def\pdens{\rho}%

\global\long\def\ms#1{\mathfrak{#1}}%

\global\long\def\Hfl{\Phi}%

\global\long\def\cud{\v j}%
\global\long\def\mag{\v M}%
\global\long\def\vp{\v A}%
\global\long\def\magf{\v B}%
\global\long\def\magi{\v H}%

\global\long\def\pfun{\mathscr{P}}%

\title[p-Form Electrodynamics as a Stress Theory]{$p$-Form Electrodynamics as a Stress Theory: Free and Bound Currents,
Gauge Invariance, and Mechanical Interactions}
\author{Vladimir Gol'dshtein, Wolfgang H.~Müller, and Reuven Segev}
\date{\today}
\begin{abstract}
We formulate electrodynamics with a $p$-form potential $\ga$ on
an oriented $n$-dimensional manifold as a theory of stresses, in
which the potentials replace the virtual velocities of continuum mechanics.
No constitutive relation, e.g., the aether relations, are used. A
current system is a collection of linear functionals on potentials,
one for each region of the manifold, the values of which are interpreted
as the potential energy of the currents in the region. Each functional
is represented by a current density and by a surface current on the
boundary of the region, and the stress theory of {[}Seg02{]}, {[}Seg23,
Ch.~10{]} provides a traction stress and a variational stress. We
do not assume that the energy is gauge invariant. We assume only weak
gauge invariance, i.e., that a gauge transformation $\ga\mapsto\ga+\gb$,
with $\gb$ closed, changes the energy through the values of $\gb$
and not through its derivatives. This is the analog of the symmetry
of the stress. We show that it holds if and only if the variational
stress has the form $\mathfrak{J}_{f}\wedge\ga+\eps\,\mathfrak{g}\wedge\dee{\ga}$,
$\eps=(-1)^{n-p-1}$, with a unique free current $(n-p)$-form $\mathfrak{J}_{f}$
and a unique dipole $(n-p-1)$-form $\mathfrak{g}$. The traction
stress is then $\ga\mapsto\mathfrak{g}\wedge\ga$, and the body force
is the total current, free plus bound. We distinguish three levels
of gauge invariance. Weak gauge invariance gives the split into free
and bound currents, invariance under the gauge transformations that
vanish near the boundary of every region is equivalent to the conservation
of the free current, and gauge invariance of the energy of every region
is equivalent to the absence of free currents, the analog of the balance
of forces. For currents embedded in matter, we derive the mechanical
forces from the energy. They consist of the Lorentz-type force on
the total current and of a gauge-dependent term, which appears only
where free currents cross the boundary of the region. Using Cartan's
formula, we show that this term is a boundary term, which we interpret
as the flux of potential energy-momentum carried by the free currents,
so that the force on matter is the gauge-invariant part. Electrostatics
and magnetostatics in $\rthree$ are the cases $n=3$, $p=0,1$. For
$n=4$, $p=1$ we recover electromagnetism in polarizable and magnetizable
matter, including the bound charge $-\nabla\cdot\v P$, the bound
current $\bdry_{t}\v P+\nabla\cross\v M$, the conservation of charge
as interior gauge invariance, and the Lorentz force density with its
time component $-\v j\cdot\v E$.
\end{abstract}

\maketitle
\tableofcontents{}

\section{Introduction}

In two companion papers, we formulated electrostatics {[}GMS26a{]}
and magnetostatics {[}GMS26b{]} in $\rthree$ as theories of stresses.
In electrostatics, the generalized velocity is the electric potential,
and in magnetostatics it is the vector potential. In both cases, a
charge or current system is a collection of linear functionals on
potentials, one for each subbody, and Cauchy's theorem splits every
such system uniquely into free charge or current and a distribution
of dipoles. Gauge invariance of the energy is not a postulate (see
Sections 4\textendash 5). It is the analog of the balance of forces,
and it singles out the systems without free charge or current. Finally,
the forces on the matter in which the currents are embedded follow
from the energy by a single differentiation.

Magnetostatics required one more ingredient, which we called weak
gauge invariance (see Section 4). It is the assumption that a gauge
transformation $\v A\mapsto\v A+\nabla\gl$ changes the energy only
through the values of $\nabla\gl$, and not through its derivatives,
and it is demonstrated to be the exact analog of the symmetry of the
stress. In addition, the magnetostatic forces derived from the energy
contain a gauge-dependent term, which appears only where free current
crosses the boundary of the subbody. We showed that this term is the
rate at which the charge carriers transport potential momentum out
of the subbody, and that the force on matter is the gauge-invariant
part.

The purpose of this paper is to show that all of the above is a special
case of a single construction. We consider a $p$-form potential $\ga$
on a smooth oriented $n$-dimensional manifold $\mathcal{X}$, representing
space or spacetime, and we apply geometric stress theory of {[}Seg02{]}
and {[}Seg23, Ch.~10{]}, with the $p$-forms as the generalized velocities.
This setting was proposed in {[}Seg16{]}, where $p$-form electrodynamics
was obtained as an example of stress theory under the assumption that
the stress is flat, i.e., that it acts on the jet of the potential
only through its value and its exterior derivative. Here, flatness
is not assumed. By Theorem~\ref{thm:decomp}, it is equivalent to
weak gauge invariance, and the unique split of the currents into free
and bound parts follows from it. The field is the $(p+1)$-form $\mathfrak{f}=\dee{\ga}$,
so that $\dee{\mathfrak{f}}=0$. Electrostatics and magnetostatics
in $\rthree$ are the cases $n=3$, $p=0$ and $p=1$. Standard electromagnetism
is the case $n=4$, $p=1$, which we treat in detail. In the general
theory, no metric is used. The constructions use only the exterior
derivative, the wedge product, the interior product and the Lie derivative,
so that the results hold on a general manifold, and they apply to
spacetime without a choice of observer.

Geometric stress theory {[}Seg02, Seg23{]} uses two notions of stress.
A \emph{traction stress} $s$ is a section of $L(W,\Lambda^{n-1}T^{*}\mathcal{X})$,
where the generalized velocities are sections of the smooth vector
bundle $W$. It acts on generalized velocity fields to produce an
$(n-1)$-form representing the flux of power. As such, it determines
the surface forces on all regions by restriction to their boundaries,
as the Cauchy stress $\gs$ determines the traction $\gs\v{\nor}$.
A \emph{variational stress} $S$ acts on the first jet $j^{1}w$ of
the generalized velocity, which consists of the values of the generalized
velocity and its first derivatives, so that the force on a region
is the integral of $S\cdot j^{1}w$ over it. In $\rthree$, the variational
stress is represented by the pair $(\v r,\gs)$ of continuum mechanics,
which acts on $(\v w,\nabla\v w)$. Every variational stress determines
a traction stress, but a traction stress determines a variational
stress only up to a term of zeroth order, which acts on the values
of $w$ alone. Both notions, and the relation between them, are reviewed
in Section~\ref{sec:prelim}.

In the analogy with continuum mechanics:
\begin{itemize}
\item the current system plays the role of a force system, with the current
density as body force and the surface current as surface traction;
\item the potential energy plays the role of the virtual power;
\item the dipole form $\mathfrak{g}$ plays the role of the traction stress,
and the free current $\mathfrak{J}_{f}$ plays the role of the unbalanced
body force;
\item gauge transformations $\ga\mapsto\ga+\gb$, for a closed form $\gb$,
play the role of rigid motions.
\end{itemize}
Under weak gauge invariance, the energy of a region $\mathcal{R}$
takes the form 
\begin{equation}
U_{\mathcal{R}}(\ga)=\int_{\mathcal{R}}\mathfrak{J}_{f}\wedge\ga+\eps\int_{\mathcal{R}}\mathfrak{g}\wedge\mathfrak{f},\qquad\eps:=(-1)^{n-p-1},\label{eq:U-intro}
\end{equation}
where $\wedge$ denotes the exterior product of forms. For $n=3$,
$p=1$, this is the magnetostatic energy $-\int(\v j_{f}\cdot\v A+\v M\cdot\v B)\dV$
of {[}GMS26b{]}, and for $n=4$, $p=1$, it is minus the interaction
term of the Lagrangian of a polarizable and magnetizable medium carrying
free charge and current, integrated over a region of spacetime.

The main results are the following.
\begin{itemize}
\item Weak gauge invariance (Assumption~\ref{ass:weak}) holds if and only
if the variational stress has the form $S\cdot j^{1}\ga=\mathfrak{J}_{f}\wedge\ga+\eps\,\mathfrak{g}\wedge\dee{\ga}$,
with a unique free current $(n-p)$-form $\mathfrak{J}_{f}$ and a
unique dipole $(n-p-1)$-form $\mathfrak{g}$ (Theorem~\ref{thm:decomp}).
The assumption holds automatically for $p=0$.
\item The traction stress is then $\ga\mapsto\mathfrak{g}\wedge\ga$, and
the body force is the total current $\mathfrak{J}=\mathfrak{J}_{f}-\dee{\mathfrak{g}}$,
free plus bound (Proposition~\ref{prop:traction}).
\item There are three levels of gauge invariance (Section~\ref{sec:levels}).
Weak gauge invariance is the analog of the symmetry of the stress.
Invariance under gauge transformations that vanish near the boundary
of every region is equivalent to the conservation of the free current,
$\dee{\mathfrak{J}_{f}}=0$. For $n=4$, $p=1$, this is the conservation
of charge, $\bdry_{t}\rho_{f}+\nabla\cdot\v j_{f}=0$. Gauge invariance
of the energy of every region is equivalent to $\mathfrak{J}_{f}=0$,
the analog of the balance of forces (Theorem~\ref{thm:main}).
\item For currents embedded in matter and convected with it, the force derived
from the energy is the Lorentz-type force on the total current, together
with a force on the surface current exposed by the dipoles and a gauge-dependent
term. The latter is a boundary term, which we interpret as the flux
of potential energy-momentum carried by the free current through the
boundary of the region, and not as a force on matter (Section~\ref{sec:mech}).
The variational stress of the mechanical force does not split invariantly
into a body force and a stress acting on the derivatives of the virtual
velocity. With a connection, such as the flat connection of $\rthree$,
it does, and the stress part is the asymmetric active mechanical stress
of {[}GMS26a, GMS26b{]}, whose skew-symmetric part is a couple density,
for example, $\v M\cross\v B$ in magnetostatics.
\item For $n=4$, $p=1$, the dipole form $\mathfrak{g}=\v P\mathbin{\lrcorner}\dV+\v M^{\flat}\wedge\dee t$
($\mathbin{\lrcorner}$ denotes the contraction of a form with a vector
field) carries the bound charge $-\nabla\cdot\v P$ and the bound
current $\bdry_{t}\v P+\nabla\cross\v M$. The force density is the
Lorentz force on the total charge and current, with time component
$-\v j\cdot\v E$ (Section~\ref{sec:st}).
\end{itemize}
The forces on polarized and magnetized matter, at rest and in motion,
have been studied extensively in continuum mechanics (see, e.g., {[}Mau88,
Chs. 4\textendash 6{]}, {[}EM90, Chs. 7\textendash 8{]}, {[}Kov00,
Chs. 17\textendash 18{]}, {[}DO14, Chs. 4 and 6{]}, and the references
there). In premetric electrodynamics {[}HO03{]}, charge conservation
and the Lorentz force density are postulated, and the excitation is
related to the field by a constitutive law. Here, the split of the
currents into free and bound parts follows from the representation
of a variational stress, the conservation of the free current is a
level of gauge invariance, and the Lorentz force density, in the form
used in {[}HO03{]}, follows from the energy once the currents are
embedded in matter.

Section~\ref{sec:prelim} recalls the parts of the stress theory
of {[}Seg02, Seg23{]} that we need and introduces current systems.
Section~\ref{sec:dtilde} treats the exterior derivative as a vector
bundle morphism on jets. Section~\ref{sec:general} contains weak
gauge invariance and the decomposition into free and bound currents.
Section~\ref{sec:gauge} characterizes gauge invariance, relates
it to the conservation of the free current, and distinguishes the
three levels of gauge invariance. Section~\ref{sec:mech} derives
the mechanical forces on matter and interprets the gauge-dependent
part. Section~\ref{sec:R3} shows how electrostatics and magnetostatics
in $\rthree$ are recovered, and Section~\ref{sec:st} treats electromagnetism
in spacetime. We close with some remarks in Section~\ref{sec:concl}.

Throughout, $\mathcal{X}$ is an oriented $n$-dimensional manifold,
$0\leq p\leq n-1$, and $\eps:=(-1)^{n-p-1}$. We denote by $w\mathbin{\lrcorner}$
the interior product with a vector field $w$, by $\mathcal{L}_{w}$
the Lie derivative, by $\wedge_{\ggm}$ the mapping $\eta\mapsto\ggm\wedge\eta$
of wedge multiplication by a form $\ggm$, defined on forms of any
degree, and by $\mathfrak{J}_{f}$, $\mathfrak{g}$ and $\mathfrak{f}$
the free current, the dipole form and the field. A region is a compact
$n$-dimensional submanifold $\mathcal{R}\subset\mathcal{X}$ with
smooth boundary, and $\bdry\mathcal{R}$ carries the induced orientation,
so that Stokes' theorem holds in the form $\int_{\mathcal{R}}\dee{\go}=\int_{\bdry\mathcal{R}}\go$.
We will often use the Leibniz rule 
\begin{equation}
\dee{(\go\wedge\psi)}=\dee{\go}\wedge\psi+(-1)^{k}\go\wedge\dee{\psi},\label{eq:leibniz}
\end{equation}
for a $k$-form $\go$. In the field theories we consider here, all
the forms involved are assumed to be smooth.

\section{Stresses on manifolds and current systems}

\label{sec:prelim}

We recall the notions and results of {[}Seg02{]} and {[}Seg23, Ch.~10{]}
that we need. In $\rthree$, with the velocity vectors as generalized
velocities, they reduce to the classical notions, for which the existence
of the stress is Cauchy's theorem {[}Nol59{]}, {[}TT60, Sect.~203,
p.~543{]}. A more detailed account of the classical case in $\rthree$,
with the same notation, is given in {[}GMS26a, Sect.~3{]} and {[}GMS26b,
Sect.~3{]}. Local coordinate representations of the geometric objects
and mappings are presented in Appendix \ref{app:coord}.

Let $\pi\colon W\to\mathcal{X}$ be a smooth vector bundle whose sections
$w$ are the generalized velocities.

\textbf{Force systems and the traction stress.} A force system is
a collection $\{F_{\mathcal{R}}\}$ of linear functionals on generalized
velocities, one for each region $\mathcal{R}$. In analogy with the
classical case, it is assumed that each $F_{\mathcal{R}}$ is represented
by a body force $b$, a section of $L(W,\Lambda^{n}T^{*}\mathcal{X})$,
and by a surface force $t_{\mathcal{R}}$ on $\bdry\mathcal{R}$,
a section of $L(W|_{\bdry\mathcal{R}},\Lambda^{n-1}T^{*}\bdry\mathcal{R})$
over $\bdry\mathcal{R}$ which depends on $\mathcal{R}$, in the form
\begin{equation}
F_{\mathcal{R}}(w)=\int_{\mathcal{R}}b\cdot w+\int_{\bdry\mathcal{R}}t_{\mathcal{R}}\cdot w.\label{eq:force}
\end{equation}
Here, $b\cdot w$ is an $n$-form on $\mathcal{R}$ and $t_{\mathcal{R}}\cdot w$
is an $(n-1)$-form on $\bdry\mathcal{R}$. Under the appropriate
assumptions of geometric stress theory (see {[}Seg23{]}), Cauchy's
theorem holds. That is, there is a section $s$ of $L(W,\Lambda^{n-1}T^{*}\mathcal{X})$,
the traction stress, such that, for every region, 
\begin{equation}
t_{\mathcal{R}}\cdot w=(s\cdot w)|_{\bdry\mathcal{R}},\label{eq:cauchy}
\end{equation}
the restriction of the $(n-1)$-form $s\cdot w$ to $\bdry\mathcal{R}$
{[}Seg23, (10.11)\textendash (10.12){]}. In $\rthree$, the $(n-1)$-form
$s\cdot w$ is the 2-form of the power flux vector field $\v{\gs}\trps\v w$,
and (\ref{eq:cauchy}) is $\v t_{\mathcal{R}}\cdot\v w=(\v{\gs}\trps\v w)\cdot\v{\nor}$.

\textbf{Variational stresses.} By Stokes' theorem, $\int_{\bdry\mathcal{R}}s\cdot w=\int_{\mathcal{R}}\dee{(s\cdot w)}$,
and (\ref{eq:force}) takes the form 
\begin{equation}
F_{\mathcal{R}}(w)=\int_{\mathcal{R}}S\cdot j^{1}w,\qquad S\cdot j^{1}w:=b\cdot w+\dee{(s\cdot w)}.\label{eq:var}
\end{equation}
The value of $\dee{(s\cdot w)}$ at a point depends on $w$ and its
first derivatives at that point, i.e., on the first jet $j^{1}w$.
Hence, $S$ is a section of the density dual $(J^{1}W)^{\star}=L(J^{1}W,\Lambda^{n}T^{*}\mathcal{X})$
of the first jet bundle, a variational stress. It represents the force
system on all regions by a single field. Its components in a chart
are given in (\ref{eq:Scoord}). Equation (\ref{eq:var}) is the principle
of virtual power {[}Seg23, (10.59){]}. In $\rthree$, $S\cdot j^{1}\v w=(\v r\cdot\v w+\gs_{ij}w_{i,j})\dV$,
with $\v r=\v b+\nabla\cdot\v{\gs}$, as in {[}GMS26b, Sect.~3{]}.
On a general manifold, however, there is no natural way of splitting
a jet into a value and a derivative, and the splitting of $S$ into
a zeroth-order part and a part acting on the derivatives is not invariant
{[}Seg23, Sect.~10.8{]}.

\textbf{The traction stress of a variational stress.} Conversely,
a variational stress determines a traction stress. The surjective
morphism 
\[
p_{s}\colon(J^{1}W)^{\star}\to L(W,\Lambda^{n-1}T^{*}\mathcal{X})
\]
 of {[}Seg23, (10.65){]} is the composition of the restriction to
the vertical subbundle $V^{1}_{0}J^{1}W:=\ker\pi^{1}_{0}$ with a
natural isomorphism (see also {[}Seg02, Sect.~VI{]}). Here, $\pi^{1}_{0}\colon J^{1}W\to W$
is the jet projection, $\pi^{1}_{0}(j^{1}w)=w$, so that the vertical
subbundle consists of the jets with vanishing value. Two facts follow
directly from this construction and the local representation {[}Seg23,
(10.62){]}.
\begin{enumerate}
\item[(P1)] $p_{s}(S)$ depends only on the action of $S$ on the vertical subbundle.
Hence, $\ker p_{s}$ consists of the variational stresses of the form
$\gb\circ\pi^{1}_{0}$, where $\gb$ is a section of $L(W,\Lambda^{n}T^{*}\mathcal{X})$.
Such variational stresses are of zeroth order, i.e., they act on the
value $w(x)$ only.
\item[(P2)] For the operator $\mathfrak{d}$ of {[}Seg23, (10.76)\textendash (10.77){]},
defined on sections $s$ of $L(W,\Lambda^{n-1}T^{*}\mathcal{X})$
by 
\begin{equation}
\mathfrak{d}(s)\cdot j^{1}w:=\dee{(s\cdot w)},\label{eq:dfrak}
\end{equation}
one has $p_{s}\circ\mathfrak{d}=\mathrm{Id}$. In a chart, $\mathfrak{d}$
and $p_{s}$ are given by (\ref{eq:dscoord}) and (\ref{eq:pscoord}).
\end{enumerate}
Consequently, the variational stresses with a given traction stress
$s$ form the affine family 
\begin{equation}
\big\{\mathfrak{d}(s)+\gb\circ\pi^{1}_{0}\;:\;\gb\text{ a section of }L(W,\Lambda^{n}T^{*}\mathcal{X})\big\}.\label{eq:affine}
\end{equation}
This is the precise sense in which the traction stress does not determine
the force system. The missing datum is the body force.

\textbf{Divergence and body force.} The divergence of a variational
stress is defined by {[}Seg23, (10.75){]} 
\begin{equation}
\mathrm{div}\,S\cdot w:=\dee{(p_{s}(S)\cdot w)}-S\cdot j^{1}w,\label{eq:div}
\end{equation}
and the body force balancing $S$ is {[}Seg23, (10.84)\textendash (10.85){]}
\begin{equation}
b=-\mathrm{div}\,S.\label{eq:body}
\end{equation}
Indeed, by (\ref{eq:var}), $S\cdot j^{1}w-\dee{(s\cdot w)}=b\cdot w$
when $s=p_{s}(S)$. In particular, $\mathrm{div}\,\mathfrak{d}(s)=0$,
so the variational stress $\mathfrak{d}(s)$ is self-equilibrated.
In a chart, the body force (\ref{eq:body}) is given by (\ref{eq:bcoord}).
The variational stress $\mathfrak{d}(s)$ assigns no body force, and
it assigns zero force to every region on whose boundary $s\cdot w$
vanishes.

\textbf{Current systems.} We now take the generalized velocities to
be $p$-forms, 
\begin{equation}
W=\Lambda^{p}T^{*}\mathcal{X},\qquad w=\ga.\label{eq:Wp}
\end{equation}
A current system is a collection $\{U_{\mathcal{R}}\}$ of linear
functionals on $p$-forms, one for each region $\mathcal{R}$. Since
we do not use the field equations, a current system is not regarded
as the source of the field, but as the object on which the field acts.
We use the term current\footnote{Continuous linear functionals on compactly supported differential
forms are called currents by de Rham {[}dR84{]}. A functional on $p$-forms
is a current of dimension $p$ and degree $n-p$, and an $(n-p)$-form
$\mathfrak{J}$ defines such a current by $\ga\mapsto\int\mathfrak{J}\wedge\ga$.
Thus, the free current $\mathfrak{J}_{f}$ is a current of degree
$n-p$ in de Rham's sense. Our terminology is similar but not identical.
A current system assigns a functional to every region, acting on forms
that need not vanish on its boundary, and each of these functionals
is represented by a variational stress, so that it is of order at
most one.} for all $p$. For $p=0$, a current is a distribution of charge,
as in the charge systems of {[}GMS26a{]}. For $n=4$, $p=1$, it consists
of charge and current, as the four-current of relativity. We interpret
$U_{\mathcal{R}}(\ga)$ as the potential energy of the currents in
$\mathcal{R}$ under the potential $\ga$. As in {[}GMS26a, GMS26b{]},
we may think of $\ga$ either as the potential itself or as a variation
of it, assuming that it can be varied while the currents are held
fixed. In analogy with mechanics, we assume that each $U_{\mathcal{R}}$
is represented in the form (\ref{eq:force}), by a current density
$b$ and a surface current $t_{\mathcal{R}}$ on $\bdry\mathcal{R}$,
and that Cauchy's theorem holds. By Lemma~\ref{lem:wedge} below,
the current density may be identified with an $(n-p)$-form, such
that $b\cdot\ga$ is the wedge product of the two. Thus, a current
system is represented by a variational stress $S$, a section of $(J^{1}(\Lambda^{p}T^{*}\mathcal{X}))^{\star}$,
in the form 
\begin{equation}
U_{\mathcal{R}}(\ga)=\int_{\mathcal{R}}S\cdot j^{1}\ga,\label{eq:US}
\end{equation}
and conversely, every variational stress defines a current system
by (\ref{eq:US}). In what follows, we use the term variational stress
for both.

In $\rthree$, the current system of electrostatics is the charge
system $\{Q_{\gO}\}$ of {[}GMS26a{]}, with $p=0$, and that of magnetostatics
is $\{-I_{\gO}\}$, minus the current system of {[}GMS26b{]}, with
$p=1$. For $n=4$, $p=1$, the region $\mathcal{R}$ is a region
of spacetime, and $U_{\mathcal{R}}$ is the integral over $\mathcal{R}$
of the interaction energy density, an action-like quantity (Section~\ref{sec:st}).

\section{The exterior derivative as a morphism on jets}

\label{sec:dtilde}

The exterior derivative is a linear first-order differential operator,
so its value at a point depends only on the first jet of the form
at that point. Hence, it corresponds to a vector bundle morphism over
$\mathcal{X}$, from the first jet bundle of $p$-forms to $(p+1)$-forms,
\begin{equation}
\tilde{d}\colon J^{1}(\Lambda^{p}T^{*}\mathcal{X})\longrightarrow\Lambda^{p+1}T^{*}\mathcal{X},\qquad\tilde{d}(j^{1}\ga):=\dee{\ga}.
\end{equation}
 In a chart, we write $\ga=\sum_{I}\ga_{I}\,\dee{x^{I}}$, with the
sum taken over increasing multi-indices $I=(i_{1},\dots,i_{p})$ and
\[
\dee{x^{I}}=\dee{x^{i_{1}}}\wedge\cdots\wedge\dee{x^{i_{p}}}.
\]
A jet $A\in J^{1}(\Lambda^{p}T^{*}\mathcal{X})$ at a point $x$ is
represented by its coordinates, the collection $(x^{i},a_{I},a_{Ik})$,
where $i,k=1,\dots,n$ and $I$ runs over the increasing multi-indices
of length $p$. For $A=j^{1}_{x}\ga$, $a_{I}=\ga_{I}(x)$ and $a_{Ik}=\bdry\ga_{I}/\bdry x^{k}(x)$.
Then 
\begin{equation}
\tilde{d}(A)=\sum_{I}\sum^{n}_{k=1}a_{Ik}\,\dee{x^{k}}\wedge\dee{x^{I}},\label{eq:dtlocal}
\end{equation}
where both indices are summed over, as they are in 
\[
\dee{\ga}=\sum_{I,k}\frac{\bdry\ga_{I}}{\bdry x^{k}}\,\dee{x^{k}}\wedge\dee{x^{I}}.
\]
This is the same kind of factorization as the one used for $\mathfrak{d}(s)$
in {[}Seg23, (10.80){]}. Equation (\ref{eq:dtlocal}) shows that $\tilde{d}$
does not see the values $a_{I}$. Its restriction to the vertical
subbundle 
\[
V:=V^{1}_{0}J^{1}(\Lambda^{p}T^{*}\mathcal{X})=\ker\pi^{1}_{0},
\]
the jets with vanishing value, is surjective onto $\Lambda^{p+1}T^{*}\mathcal{X}$,
since every basis element $\dee{x^{k}}\wedge\dee{x^{I}}$, with $k\notin I$,
is the image of the vertical jet whose only nonzero component is $a_{Ik}=1$.
Coordinate representations of the other objects used in this paper
are collected in Appendix~\ref{app:coord}.

The kernel of $\tilde{d}$ will play the role of the jets of rigid
velocity fields.
\begin{lem}
\label{lem:kernel} Let $x_{0}\in\mathcal{X}$. Every element of $\ker\tilde{d}_{x_{0}}$
is the first jet at $x_{0}$ of a closed $p$-form defined in a neighborhood
of $x_{0}$, and every element of $\ker\tilde{d}_{x_{0}}\cap V_{x_{0}}$
is the first jet at $x_{0}$ of a closed $p$-form $\ggm$ that vanishes
at $x_{0}$. For $p\geq1$, the closed form $\ggm$ may be taken to
be exact, so that $\ggm=\dee{\gl}$, where $\gl$ is a $(p-1)$-form
defined in a contractible neighborhood of $x_{0}$. For $p=0$, it
is a constant function, and $\ker\tilde{d}_{x_{0}}\cap V_{x_{0}}=\{0\}$.
\end{lem}

\begin{proof}
Let $A\in\ker\tilde{d}_{x_{0}}$ have the coordinates $(x^{i}_{0},a_{I},a_{Ik})$
in a chart, so that $\sum_{I,k}a_{Ik}\,\dee{x^{k}}\wedge\dee{x^{I}}=0$.
The form with affine coefficients 
\[
\hat{\ga}:=\sum_{I}\Bigl(a_{I}+\sum_{k}a_{Ik}(x^{k}-x^{k}_{0})\Bigr)\dee{x^{I}}
\]
has the given first jet at $x_{0}$. Its exterior derivative, 
\[
\dee{\hat{\ga}}=\sum_{I,k}a_{Ik}\,\dee{x^{k}}\wedge\dee{x^{I}},
\]
has constant coefficients, and it vanishes identically on the chart
domain. If the jet is vertical, then $a_{I}=0$ and $\hat{\ga}(x_{0})=0$.
Restricting $\hat{\ga}$ to a ball about $x_{0}$, it is exact for
$p\geq1$, by the Poincaré lemma. For $p=0$, there is a single component
$a$, with derivatives $a_{k}$, and the condition is $a_{k}=0$ for
all $k$, so $\hat{\ga}=a$ is constant, and it vanishes at $x_{0}$
only if the jet is zero.
\end{proof}

\begin{lem}
\label{lem:wedge} For $0\leq k\leq n$, the mapping $h\mapsto(\vph\mapsto h\wedge\vph)$
is a natural vector bundle isomorphism $\Lambda^{n-k}T^{*}\mathcal{X}\to L(\Lambda^{k}T^{*}\mathcal{X},\Lambda^{n}T^{*}\mathcal{X})$.
\end{lem}

\begin{proof}
At each point, the wedge product $\Lambda^{n-k}T^{*}_{x}\mathcal{X}\times\Lambda^{k}T^{*}_{x}\mathcal{X}\to\Lambda^{n}T^{*}_{x}\mathcal{X}$
is a nondegenerate pairing. Indeed, in a basis, $\dee{x^{J}}\wedge\dee{x^{I}}=\pm\dee{x^{1}}\wedge\cdots\wedge\dee{x^{n}}$
if $J$ is the complement of $I$ and vanishes otherwise. The two
spaces have the same dimension $\binom{n}{k}$, so the mapping is
an isomorphism at each point, and it is clearly smooth and independent
of any choices.
\end{proof}

Below, we will use the lemma with $k=p+1$, for the dipole form, with
$k=p$, for the free current, and with $k=p-1$, for the conservation
of the free current.

\section{Free and bound currents}

\label{sec:general}

\subsection{Gauge transformations and weak gauge invariance}

A gauge transformation is the replacement of $\ga$ by $\ga+\gb$,
where $\gb$ is a closed $p$-form. For $p\geq1$, a closed form is
locally exact, $\gb=\dee{\gl}$ for a $(p-1)$-form $\gl$, and for
$p=0$, $\gb$ is a constant on every connected component. We call
$\gb$ a gauge form. The field $\mathfrak{f}=\dee{\ga}$ is obviously
independent of $\gb$. For a variational stress $S$, the energy of
the current in a region $\mathcal{R}$ changes by 
\begin{equation}
U_{\mathcal{R}}(\ga+\gb)-U_{\mathcal{R}}(\ga)=U_{\mathcal{R}}(\gb)=\int_{\mathcal{R}}S\cdot j^{1}\gb.\label{eq:gaugechange}
\end{equation}
For $p=0$, $\gb=c$ is a constant, and the change is $c$ times the
net charge of $\mathcal{R}$, a physical effect {[}GMS26a{]}. For
$n=3$, $p=1$, a single current element $I\dee{\v l}$ at $\v y$
has the energy $-I\dee{\v l}\cdot\v A(\v y)$, which changes by $-I\dee{\v l}\cdot\nabla\gl(\v y)$
{[}GMS26b{]}. Hence, we do not assume that the energy is gauge invariant.

We now consider the first jets of the gauge forms at a point. By Lemma~\ref{lem:kernel},
the first jets of gauge forms at $x$ are exactly the elements of
$\ker\tilde{d}_{x}$. Their values can be prescribed arbitrarily,
since every $p$-form $a$ at $x$ is the value of the closed form
with constant coefficients $a$ in a chart. The jets of the gauge
forms that vanish at $x$ constitute $\ker\tilde{d}_{x}\cap V_{x}$.
In $\rthree$, with $p=1$, the gauge forms are the fields $\nabla\gl$.
At a point, the value $\nabla\gl$ can be any vector, and, independently,
the gradient $\nabla\nabla\gl$ can be any symmetric matrix. They
take the place of the rigid velocity fields $\v c+\v{\go}\cross\v x$
of continuum mechanics, whose values are arbitrary and whose gradients
are arbitrary skew-symmetric matrices. In general, the jets of the
gauge forms that vanish at a point, $\ker\tilde{d}\cap V$, play the
role of the gradients of rigid fields, i.e., of the rotations about
the point.

In continuum mechanics, the stress is symmetric if and only if the
power of the rigid velocity fields depends only on their values {[}GMS26b,
Prop.~3.1{]}. Equivalently, the power density at each point is invariant
under the rotations about that point. We make the analogous assumption
here. It is equivalent to the assumption of {[}Seg16, Sect.~7{]}
that the stress is flat (see Section~\ref{sec:flat} below).
\begin{assumption}[Weak gauge invariance]
\label{ass:weak} The change of the energy under a gauge transformation
depends only on the values of the gauge form. That is, there is a
section $R$ of $L(\Lambda^{p}T^{*}\mathcal{X},\Lambda^{n}T^{*}\mathcal{X})$
such that 
\begin{equation}
S\cdot j^{1}\gb=R(\gb)\label{eq:weak}
\end{equation}
for every gauge form $\gb$, defined on any open subset of $\mathcal{X}$.
A variational stress, or a current system, that satisfies this assumption
is called weakly gauge invariant.
\end{assumption}

Since regions can be taken arbitrarily small, (\ref{eq:weak}) is
equivalent to $U_{\mathcal{R}}(\gb)=\int_{\mathcal{R}}R(\gb)$ for
every region $\mathcal{R}$ and every gauge form $\gb$ defined in
a neighborhood of $\mathcal{R}$. Thus, the energy depends on the
values of the gauge form itself, through a pairing with a current,
but not its derivatives. The term weak gauge invariance is motivated
by Theorem~\ref{thm:decomp}, which shows that the assumption holds
if and only if the energy density at every point $x$ is invariant
under every gauge transformation whose gauge form vanishes at $x$,
though not, in general, under the others. In $\rthree$, with $p=1$,
it is equivalent to the skew-symmetry of the analog $\v{\Sigma}$
of the traction stress of magnetostatics {[}GMS26b, Prop.~4.2{]}.

\subsection{The decomposition}
\begin{thm}
\label{thm:decomp} For a variational stress $S$, a section of $(J^{1}(\Lambda^{p}T^{*}\mathcal{X}))^{\star}$,
the following are equivalent.
\begin{enumerate}
\item[(i)] $S$ is weakly gauge invariant.
\item[(ii)] $S$ annihilates $\ker\tilde{d}\cap V$. That is, $S\cdot j^{1}_{x}\gb=0$
for every point $x$ and every gauge form $\gb$ with $\gb(x)=0$.
\item[(iii)] There are unique sections $R$ of $L(\Lambda^{p}T^{*}\mathcal{X},\Lambda^{n}T^{*}\mathcal{X})$
and $\hat{S}$ of $L(\Lambda^{p+1}T^{*}\mathcal{X},\Lambda^{n}T^{*}\mathcal{X})$
such that 
\begin{equation}
S=R\circ\pi^{1}_{0}+\hat{S}\circ\tilde{d}.\label{eq:RShat}
\end{equation}
\item[(iv)] There are a unique $(n-p)$-form $\mathfrak{J}_{f}$ and a unique
$(n-p-1)$-form $\mathfrak{g}$ such that 
\begin{equation}
S\cdot j^{1}\ga=\mathfrak{J}_{f}\wedge\ga+\eps\,\mathfrak{g}\wedge\dee{\ga}\qquad\text{for every \ensuremath{p}-form }\ga.\label{eq:SJg}
\end{equation}
\end{enumerate}
For $p=0$, the conditions hold for every variational stress.
\end{thm}

\begin{proof}
(i)$\Rightarrow$(ii). If $\gb$ is a gauge form and $\gb(x)=0$,
then, by (\ref{eq:weak}), $S\cdot j^{1}_{x}\gb=R(\gb(x))=0$. By
Lemma~\ref{lem:kernel}, the jets of such gauge forms exhaust $\ker\tilde{d}_{x}\cap V_{x}$.

(ii)$\Rightarrow$(iii). Fix a point $x$. The restriction $\tilde{d}_{x}|_{V_{x}}\colon V_{x}\to\Lambda^{p+1}T^{*}_{x}\mathcal{X}$
is surjective, and, by (ii), the restriction of $S_{x}$ to $V_{x}$
vanishes on its kernel. A linear mapping $L$ that vanishes on the
kernel of a surjective linear mapping $T$ factors uniquely through
it.\footnote{In detail, let $T\colon V_{1}\to V_{2}$ be a surjective linear mapping
and $L\colon V_{1}\to V_{3}$ a linear mapping with $\ker T\subset\ker L$.
For $z\in V_{2}$, choose $v\in V_{1}$ with $T(v)=z$, which is possible
since $T$ is surjective, and set $\hat{L}(z):=L(v)$. If also $T(v')=z$,
then $v-v'\in\ker T\subset\ker L$, so that $L(v')=L(v)$, and $\hat{L}$
is well defined. It is linear, since, if $T(v)=z$ and $T(v')=z'$,
then $T(av+bv')=az+bz'$, so that $\hat{L}(az+bz')=L(av+bv')=a\hat{L}(z)+b\hat{L}(z')$.
By construction, $\hat{L}\circ T=L$, and $\hat{L}$ is the only mapping
with this property, since every $z$ is of the form $T(v)$.} We apply this to $T=\tilde{d}_{x}|_{V_{x}}$ and $L=S_{x}|_{V_{x}}$.
Hence, there is a unique linear mapping $\hat{S}_{x}\colon\Lambda^{p+1}T^{*}_{x}\mathcal{X}\to\Lambda^{n}T^{*}_{x}\mathcal{X}$
with $S_{x}|_{V_{x}}=\hat{S}_{x}\circ\tilde{d}_{x}|_{V_{x}}$. The
difference $S_{x}-\hat{S}_{x}\circ\tilde{d}_{x}$ vanishes on $V_{x}=\ker(\pi^{1}_{0})_{x}$,
and $(\pi^{1}_{0})_{x}$ is surjective, so the difference equals $R_{x}\circ(\pi^{1}_{0})_{x}$
for a unique $R_{x}$. As for uniqueness, if (\ref{eq:RShat}) holds,
then, since $\pi^{1}_{0}$ vanishes on $V$, $\hat{S}$ is determined
by the restriction of $S$ to $V$, and then $R$ is determined by
$S-\hat{S}\circ\tilde{d}$. Finally, in a chart, the components of
$\hat{S}$ are among the components of $S$ that act on the derivatives
$a_{Ik}$, by (\ref{eq:dtlocal}), and the components of $R$ are
the components of $S$ that act on the values $a_{I}$ (see Appendix~\ref{app:coord}).
Thus, $\hat{S}$ and $R$ are smooth.

(iii)$\Rightarrow$(i). For a gauge form $\gb$, $\tilde{d}(j^{1}\gb)=\dee{\gb}=0$,
so that $S\cdot j^{1}\gb=R(\gb)$.

(iii)$\Leftrightarrow$(iv). By Lemma~\ref{lem:wedge}, with $k=p$
and $k=p+1$, there are unique forms $\mathfrak{J}_{f}$ and $\mathfrak{g}$
with $R(\vph)=\mathfrak{J}_{f}\wedge\vph$ and $\hat{S}(\psi)=\eps\,\mathfrak{g}\wedge\psi$.
Then (\ref{eq:RShat}) is (\ref{eq:SJg}).

For $p=0$, $\ker\tilde{d}\cap V=\{0\}$, by Lemma~\ref{lem:kernel},
so (ii) holds trivially.
\end{proof}

In a chart, condition (iii) is (\ref{eq:weakcoord}) of Appendix~\ref{app:coord},
i.e., the component $S^{Ik}$ depends only on the set $\{k\}\cup I$,
up to sign. It follows that the energy of the current in a region
is given by 
\begin{equation}
U_{\mathcal{R}}(\ga)=\int_{\mathcal{R}}\mathfrak{J}_{f}\wedge\ga+\eps\int_{\mathcal{R}}\mathfrak{g}\wedge\mathfrak{f},\qquad\mathfrak{f}=\dee{\ga}.\label{eq:U}
\end{equation}
Under a gauge transformation given by a form $\gb$, since $\dee{\gb}=0$,
the change in energy is given by
\begin{equation}
U_{\mathcal{R}}(\ga+\gb)-U_{\mathcal{R}}(\ga)=\int_{\mathcal{R}}\mathfrak{J}_{f}\wedge\gb.\label{eq:gaugechange2}
\end{equation}
We call $\mathfrak{J}_{f}$ the free current and $\mathfrak{g}$ the
dipole form. The free current is conjugate to the potential, and the
dipole form is conjugate to the field $\mathfrak{f}$. The pair $(\mathfrak{J}_{f},\mathfrak{g})$
is determined by $S$, and no relation between the two is assumed.
The sign $\eps$ is chosen so that the traction stress takes the simple
form (\ref{eq:sg}) below. In $\rthree$, with $p=0$, $\mathfrak{J}_{f}$
is the 3-form of the free charge density and $\mathfrak{g}$ is the
2-form of the polarization. For $p=1$, $\mathfrak{J}_{f}$ is the
2-form of minus the free current density and $\mathfrak{g}$ is the
1-form of the magnetization (Section~\ref{sec:R3}).

The decomposition (\ref{eq:RShat}) deserves a comment. In {[}Seg23,
Sect.~10.8{]}, it is noted that the zeroth-order components of a
general variational stress cannot be isolated invariantly, and {[}Seg23,
Sect.~10.9{]} introduces a connection in order to split the jet.
For $W=\Lambda^{p}T^{*}\mathcal{X}$ and a weakly gauge-invariant
variational stress, no connection is needed. The exterior derivative
furnishes a natural substitute, and the decomposition (\ref{eq:RShat})
is invariant. The same holds for $\mathfrak{J}_{f}$ and $\mathfrak{g}$.
\begin{rem}[Counting components]
At each point, a general variational stress has $\binom{n}{p}+n\binom{n}{p}=(n+1)\binom{n}{p}$
components, acting on the values $\ga_{I}$ and on the derivatives
$\ga_{I,k}$. The forms $\mathfrak{J}_{f}$ and $\mathfrak{g}$ have
$\binom{n}{p}+\binom{n}{p+1}$ components. The two numbers agree for
$p=0$, where weak gauge invariance holds trivially, and differ for
$p\geq1$. For $n=3$, $p=1$, the count is $12$ against $3+3$.
The six missing components are those acting on the symmetric part
of $\nabla\v A$, i.e., the symmetric part of $\v{\Sigma}$ in {[}GMS26b{]}.
For $n=4$, $p=1$, the count is $20$ against $4+6$.
\end{rem}

\subsection{Weak gauge invariance and flat forces}

\label{sec:flat}

The representation (\ref{eq:RShat}) appeared in {[}Seg16, Sect.~7(c){]},
where a variational stress of the form 
\begin{equation}
S\cdot j^{1}\ga=S_{0}\cdot\ga+S_{1}\cdot\dee{\ga},\label{eq:flatS}
\end{equation}
which acts on the jet of $\ga$ only through the value $\ga$ and
the exterior derivative $\dee{\ga}$, is called flat. In the terminology
of Theorem~\ref{thm:decomp}, $S_{0}=R$ and $S_{1}=\hat{S}$. Thus,
Theorem~\ref{thm:decomp} shows that a variational stress is flat
if and only if it is weakly gauge invariant. The assumption that the
stress acts only on the exterior derivative of the jet, which is the
basic assumption leading to $p$-form electrodynamics in {[}Seg16,
Sect.~7{]}, is therefore the exact analog of the symmetry of the
stress, and it has a physical meaning in terms of the dependence of
the energy on gauge transformations. For magnetostatics, it was noted
in {[}Seg16, Sect.~1(a){]} that a skew-symmetric stress leads to
the equations of magnetostatics.

In {[}Seg16, Sect.~7(c){]}, flatness is also related to continuity.
Consider the space of compactly supported continuous $p$-forms whose
exterior derivatives are continuous, with the flat topology, in which
a sequence converges to zero if the supports lie in a common compact
set and both the forms and their exterior derivatives converge to
zero uniformly {[}Seg16, (7.22){]}. The terminology follows the flat
chains of Whitney and Federer {[}Whi57{]}, {[}Fed69{]}. It is shown
in {[}Seg16, Sect.~7(c){]} that a linear functional $F$ on $p$-forms
is continuous with respect to the flat topology if and only if it
is of the form 
\begin{equation}
F(\ga)=H(\ga)+G(\dee{\ga}),\label{eq:flatF}
\end{equation}
where $H$ and $G$ are de Rham currents of order zero, i.e., they
are represented by measures. For a weakly gauge-invariant current
system with compact support, the total energy 
\[
U_{\mathcal{X}}(\ga)=\int_{\mathcal{X}}\mathfrak{J}_{f}\wedge\ga+\eps\int_{\mathcal{X}}\mathfrak{g}\wedge\dee{\ga}
\]
is of this form. It follows that weak gauge invariance may be stated
as a continuity requirement. The energy depends continuously on the
potential and on the field, and it does not depend on the other derivatives
of the potential. This formulation also applies to singular currents,
such as point charges, current filaments and dipole layers, whose
energies are represented by measures.

\subsection{The electrodynamics analogs of stress and body force}
\begin{prop}
\label{prop:traction} Let $S$ be weakly gauge invariant, given by
(\ref{eq:SJg}), and define the bound current and the total current
by 
\begin{equation}
\mathfrak{J}_{b}:=-\dee{\mathfrak{g}},\qquad\mathfrak{J}:=\mathfrak{J}_{f}+\mathfrak{J}_{b}=\mathfrak{J}_{f}-\dee{\mathfrak{g}}.\label{eq:Jb}
\end{equation}
Then the traction stress of $S$ is given by 
\begin{equation}
p_{s}(S)=\wedge_{\mathfrak{g}},\qquad\text{i.e.,}\qquad p_{s}(S)\cdot\ga=\mathfrak{g}\wedge\ga.\label{eq:sg}
\end{equation}
The variational stress decomposes in the form 
\begin{equation}
S=\mathfrak{d}(\wedge_{\mathfrak{g}})+\wedge_{\mathfrak{J}}\circ\pi^{1}_{0},\label{eq:Sdecomp}
\end{equation}
and the body force is the total current, 
\begin{equation}
b=-\mathrm{div}\,S=\mathfrak{J}.\label{eq:divS}
\end{equation}
Consequently, the principle of virtual power takes the form 
\begin{equation}
U_{\mathcal{R}}(\ga)=\int_{\mathcal{R}}\mathfrak{J}\wedge\ga+\int_{\bdry\mathcal{R}}\mathfrak{g}\wedge\ga=\int_{\mathcal{R}}\mathfrak{J}_{f}\wedge\ga+\eps\int_{\mathcal{R}}\mathfrak{g}\wedge\dee{\ga}.\label{eq:pvpnew}
\end{equation}
\end{prop}

\begin{proof}
The form $\mathfrak{g}$ is of degree $n-p-1$, so, by the Leibniz
rule (\ref{eq:leibniz}), 
\begin{equation}
\dee{(\mathfrak{g}\wedge\ga)}=\dee{\mathfrak{g}}\wedge\ga+\eps\,\mathfrak{g}\wedge\dee{\ga}.\label{eq:dgalpha}
\end{equation}
Substituting $\eps\,\mathfrak{g}\wedge\dee{\ga}=\dee{(\mathfrak{g}\wedge\ga)}-\dee{\mathfrak{g}}\wedge\ga$
in (\ref{eq:SJg}), 
\[
S\cdot j^{1}\ga=\dee{(\mathfrak{g}\wedge\ga)}+(\mathfrak{J}_{f}-\dee{\mathfrak{g}})\wedge\ga=\mathfrak{d}(\wedge_{\mathfrak{g}})\cdot j^{1}\ga+\wedge_{\mathfrak{J}}\big(\pi^{1}_{0}(j^{1}\ga)\big),
\]
by the definition (\ref{eq:dfrak}) of $\mathfrak{d}$. This is (\ref{eq:Sdecomp}).
The second term is of zeroth order. Hence, by (P1) and (P2), $p_{s}(S)=p_{s}(\mathfrak{d}(\wedge_{\mathfrak{g}}))=\wedge_{\mathfrak{g}}$,
which is (\ref{eq:sg}). By (\ref{eq:div}), 
\[
\mathrm{div}\,S\cdot\ga=\dee{(\mathfrak{g}\wedge\ga)}-S\cdot j^{1}\ga=-\mathfrak{J}\wedge\ga,
\]
and (\ref{eq:divS}) follows from (\ref{eq:body}). Finally, (\ref{eq:pvpnew})
follows by integrating (\ref{eq:Sdecomp}) over $\mathcal{R}$ and
using Stokes' theorem for the first term.
\end{proof}

\begin{cor}
\label{cor:form} The traction stress of a weakly gauge-invariant
variational stress is of the form $\wedge_{\mathfrak{g}}$ for an
$(n-p-1)$-form $\mathfrak{g}$.
\end{cor}

The free current, being of zeroth order, does not contribute to the
traction stress. By Lemma~\ref{lem:wedge}, the traction stresses
of this form constitute a subbundle of $L(\Lambda^{p}T^{*}\mathcal{X},\Lambda^{n-1}T^{*}\mathcal{X})$
of rank $\binom{n}{p+1}$, whereas the full bundle has rank $n\binom{n}{p}$.
Hence, except for $p=0$, weak gauge invariance is a genuine restriction
on the traction stress. For $n=3$ (Table~\ref{tab:rank}), it is
no restriction for $p=0$, it gives the skew-symmetric traction stress
of magnetostatics for $p=1$, and a multiple of the identity for $p=2$.
For $n=4$, $p=1$, the rank is $6$ out of $16$. This is the analog
of the symmetry of the stress in continuum mechanics, which reduces
the nine components of the traction stress to six.

\begin{table}[ht]
\begin{tabular}{cccl}
\toprule 
$p$ & full rank & rank of the subbundle & traction stress\tabularnewline
\midrule 
0 & 3 & 3 & every traction stress ($\v P$)\tabularnewline
1 & 9 & 3 & skew-symmetric ($\v M$)\tabularnewline
2 & 9 & 1 & multiple of the identity\tabularnewline
\bottomrule
\end{tabular}\vspace{3mm}
 \caption{Traction stresses of weakly gauge-invariant variational stresses for
$n=3$. The full rank is that of $L(\Lambda^{p}T^{*}\mathcal{X},\Lambda^{2}T^{*}\mathcal{X})$.}
\label{tab:rank}
\end{table}

\textbf{Interpretation.} By (\ref{eq:pvpnew}), the energy of the
currents in $\mathcal{R}$ is the energy of the total current $\mathfrak{J}$
in $\mathcal{R}$, plus the energy of the surface current $\mathfrak{g}|_{\bdry\mathcal{R}}$
exposed by the dipoles on $\bdry\mathcal{R}$. A distribution of dipoles
with density $\mathfrak{g}$ carries the bound current $\mathfrak{J}_{b}=-\dee{\mathfrak{g}}$
in the interior of $\mathcal{R}$ and the surface current $\mathfrak{g}$
on its boundary. For $n=3$, $p=0$, these are the bound charge $-\nabla\cdot\v P$
and the bound surface charge $\v P\cdot\v{\nor}$. For $p=1$, they
are the bound current $\nabla\cross\v M$ and the bound surface current
$\v M\cross\v{\nor}$ (Section~\ref{sec:R3}). The surface current
is defined for every region, not only for the boundary of the medium.
On an internal surface, it is the current exposed when the region
is cut out, i.e., the halves of the dipoles severed by the cut. This
is the analog of the traction on an internal surface in continuum
mechanics. The bound current is automatically conserved, $\dee{\mathfrak{J}_{b}}=-\dee{\dee{\mathfrak{g}}}=0$.

Over all of $\mathcal{X}$, and for a potential of compact support,
the boundary term of (\ref{eq:pvpnew}) drops out, and $U_{\mathcal{X}}(\ga)=\int_{\mathcal{X}}\mathfrak{J}\wedge\ga$
is the energy of the total current. For a region, the energy is not
a boundary functional, i.e., it is not the integral of an exact form,
unless $\mathfrak{J}=0$.

\section{Gauge invariance}

\label{sec:gauge}

\subsection{The main result}

By (\ref{eq:gaugechange2}), the energy of a weakly gauge-invariant
current system changes under a gauge transformation by $\int_{\mathcal{R}}\mathfrak{J}_{f}\wedge\gb$,
and this change is physical, as for the current element of Section~\ref{sec:general}.
Thus, gauge invariance is a property of a special class of current
systems, just as equilibrium is a property of a special class of force
systems.
\begin{defn}
\label{def:gauge} A variational stress $S$, or the current system
it represents, is gauge invariant if $U_{\mathcal{R}}(\ga+\gb)=U_{\mathcal{R}}(\ga)$
for every region $\mathcal{R}$, every $p$-form $\ga$, and every
gauge form $\gb$ defined in a neighborhood of $\mathcal{R}$. Equivalently,
$S\cdot j^{1}\gb=0$ for every gauge form $\gb$, defined on any open
set.
\end{defn}

\begin{thm}
\label{thm:main} For a variational stress $S$, the following are
equivalent.
\begin{enumerate}
\item[(i)] $S$ is gauge invariant.
\item[(ii)] $S$ annihilates $\ker\tilde{d}$.
\item[(iii)] $S=\hat{S}\circ\tilde{d}$ for a unique section $\hat{S}$ of $L(\Lambda^{p+1}T^{*}\mathcal{X},\Lambda^{n}T^{*}\mathcal{X})$.
\item[(iv)] $S$ is weakly gauge invariant and $\mathfrak{J}_{f}=0$. That is,
$S\cdot j^{1}\ga=\eps\,\mathfrak{g}\wedge\dee{\ga}$ for a unique
$(n-p-1)$-form $\mathfrak{g}$, and 
\begin{equation}
U_{\mathcal{R}}(\ga)=\eps\int_{\mathcal{R}}\mathfrak{g}\wedge\mathfrak{f}\label{eq:Ugauge}
\end{equation}
for every region $\mathcal{R}$.
\end{enumerate}
\end{thm}

\begin{proof}
(i)$\Leftrightarrow$(ii). By Lemma~\ref{lem:kernel}, $\ker\tilde{d}_{x}$
consists exactly of the first jets at $x$ of gauge forms. (ii)$\Leftrightarrow$(iii).
$\tilde{d}_{x}$ is surjective, so a linear mapping vanishing on $\ker\tilde{d}_{x}$
factors uniquely through it, and smoothness follows as in the proof
of Theorem~\ref{thm:decomp}. (iii)$\Rightarrow$(iv). Condition
(iii) is condition (iii) of Theorem~\ref{thm:decomp} with $R=0$,
and, by Lemma~\ref{lem:wedge}, $R=0$ if and only if $\mathfrak{J}_{f}=0$.
(iv)$\Rightarrow$(i). If $\mathfrak{J}_{f}=0$, then, by (\ref{eq:gaugechange2}),
the energy does not change under gauge transformations.
\end{proof}

Note that Theorem~\ref{thm:main} does not assume weak gauge invariance.
Gauge invariance implies it, just as invariance of the power under
all rigid motions implies both the balance of forces and the symmetry
of the stress {[}GMS26b, Prop.~3.1(b){]}. We call the property of
Definition~\ref{def:gauge} simply gauge invariance, as in {[}GMS26a,
GMS26b{]}. For $p\geq1$, the proof uses only exact gauge forms $\dee{\gl}$,
whose first jets already exhaust $\ker\tilde{d}$ by Lemma~\ref{lem:kernel}.
Thus, it makes no difference whether, in Definition~\ref{def:gauge},
one allows all closed forms or only exact ones.

We conclude that gauge invariance of the energy is not a postulate.
It is a characterization of the current systems without free current,
those of pure dipole distributions. It is to $p$-form electrodynamics
what the balance of forces on every subbody is to continuum mechanics,
which characterizes the force systems that are represented by a stress
acting on the velocity gradient alone. A body carrying a free current
is the analog of a body acted upon by unbalanced forces. The representation
(\ref{eq:Ugauge}) of the energy is the analog of the representation
$\int\gs_{ij}w_{i,j}\dV$ of the power of an equilibrated force system.
\begin{prop}
\label{prop:unique} The variational stresses with a given traction
stress $\wedge_{\mathfrak{g}}$ are all weakly gauge invariant, and
they are parametrized by their free current $\mathfrak{J}_{f}$. Exactly
one of them is gauge invariant, namely $\eps\,\wedge_{\mathfrak{g}}\circ\tilde{d}$.
\end{prop}

\begin{proof}
By (\ref{eq:affine}), the variational stresses with traction stress
$\wedge_{\mathfrak{g}}$ are $\mathfrak{d}(\wedge_{\mathfrak{g}})+\gb\circ\pi^{1}_{0}$,
with $\gb$ an arbitrary section of $L(\Lambda^{p}T^{*}\mathcal{X},\Lambda^{n}T^{*}\mathcal{X})$.
By Lemma~\ref{lem:wedge}, $\gb=\wedge_{\mathfrak{K}}$ for a unique
$(n-p)$-form $\mathfrak{K}$. By (\ref{eq:dgalpha}), the variational
stress acts in the form 
\[
\dee{(\mathfrak{g}\wedge\ga)}+\mathfrak{K}\wedge\ga=(\mathfrak{K}+\dee{\mathfrak{g}})\wedge\ga+\eps\,\mathfrak{g}\wedge\dee{\ga},
\]
which is (\ref{eq:SJg}) with $\mathfrak{J}_{f}=\mathfrak{K}+\dee{\mathfrak{g}}$.
Hence, it is weakly gauge invariant, and $\mathfrak{K}\mapsto\mathfrak{J}_{f}$
is a bijection. By Theorem~\ref{thm:main}, the gauge-invariant one
has $\mathfrak{J}_{f}=0$.
\end{proof}

In general, the traction stress determines only the dipole part, and
the free current is the additional datum needed to determine the energy,
exactly as the unbalanced body force is in continuum mechanics. When
there is no free current, the traction stress determines the energy.

\subsection{Consequences}

\textbf{The energy depends only on the field.} For a current system
without free current, the energy (\ref{eq:Ugauge}) depends on $\ga$
only through $\mathfrak{f}$. In general, it also depends on $\ga$
itself, through the free current.

\textbf{Gauge invariance of the energy of a single region.} The energy
of a single region $\mathcal{R}$ may be gauge invariant even if $\mathfrak{J}_{f}\neq0$.
For $p=0$, the gauge forms are constants $c$, and, by (\ref{eq:gaugechange2}),
the energy of $\mathcal{R}$ is gauge invariant if and only if the
net free charge of each connected component of $\mathcal{R}$ vanishes,
since a gauge form may take different constant values on different
components. For $p\geq1$, let $\gb=\dee{\gl}$, with $\gl$ a $(p-1)$-form.
Since $\mathfrak{J}_{f}$ is of degree $n-p$, the Leibniz rule (\ref{eq:leibniz})
gives 
\[
\dee{(\mathfrak{J}_{f}\wedge\gl)}=\dee{\mathfrak{J}_{f}}\wedge\gl+(-1)^{n-p}\mathfrak{J}_{f}\wedge\dee{\gl},
\]
and, since $(-1)^{n-p}=-\eps$, we get $\mathfrak{J}_{f}\wedge\dee{\gl}=-\eps\,\dee{(\mathfrak{J}_{f}\wedge\gl)}+\eps\,\dee{\mathfrak{J}_{f}}\wedge\gl$.
Integrating over $\mathcal{R}$ and using Stokes' theorem, 
\begin{equation}
\int_{\mathcal{R}}\mathfrak{J}_{f}\wedge\dee{\gl}=-\eps\int_{\bdry\mathcal{R}}\mathfrak{J}_{f}\wedge\gl+\eps\int_{\mathcal{R}}\dee{\mathfrak{J}_{f}}\wedge\gl.\label{eq:gaugebody}
\end{equation}
This vanishes for every $\gl$ if and only if $\dee{\mathfrak{J}_{f}}=0$
in $\mathcal{R}$ and the restriction of $\mathfrak{J}_{f}$ to $\bdry\mathcal{R}$
vanishes. Indeed, the two conditions clearly imply that both integrals
vanish. Conversely, assume that the right-hand side of (\ref{eq:gaugebody})
vanishes for every $\gl$. We first take $\gl$ that vanishes in a
neighborhood of $\bdry\mathcal{R}$. Then the boundary integral vanishes,
and $\int_{\mathcal{R}}\dee{\mathfrak{J}_{f}}\wedge\gl=0$ for all
such $\gl$. If $\dee{\mathfrak{J}_{f}}\neq0$ at some interior point
$x$, then, by Lemma~\ref{lem:wedge} with $k=p-1$, there is a $(p-1)$-form
$\gl_{0}$ at $x$ with $\dee{\mathfrak{J}_{f}}\wedge\gl_{0}\neq0$.
Multiplying the constant-coefficient extension of $\gl_{0}$ in a
chart by a nonnegative function supported in a small ball about $x$
gives an integrand that does not change sign and does not vanish identically,
a contradiction. Hence, $\dee{\mathfrak{J}_{f}}=0$ in the interior
of $\mathcal{R}$, and, by continuity, on all of $\mathcal{R}$. It
follows that $\int_{\bdry\mathcal{R}}\mathfrak{J}_{f}\wedge\gl=0$
for every $\gl$. On $\bdry\mathcal{R}$, the integrand is the wedge
product of the restrictions of $\mathfrak{J}_{f}$ and $\gl$ to $\bdry\mathcal{R}$,
an $(n-p)$-form and a $(p-1)$-form on the $(n-1)$-dimensional manifold
$\bdry\mathcal{R}$, and every $(p-1)$-form on $\bdry\mathcal{R}$
is the restriction of a $(p-1)$-form $\gl$ defined on $\mathcal{R}$.
The same argument, with Lemma~\ref{lem:wedge} applied on $\bdry\mathcal{R}$,
shows that the restriction of $\mathfrak{J}_{f}$ to $\bdry\mathcal{R}$
vanishes. We say in the latter case, $\mathfrak{J}_{f}\resto{\bdry\mathcal{R}}=0$,
that the free current does not cross $\bdry\mathcal{R}$. For $n=3$,
$p=1$, the two conditions are $\nabla\cdot\v j_{f}=0$ and $\v j_{f}\cdot\v{\nor}=0$,
i.e., the free current closes within $\mathcal{R}$ {[}GMS26b, Sect.~5.2{]}.
A region $\mathcal{R}'\subset\mathcal{R}$ through whose boundary
the free current flows does not have a gauge-invariant energy. This
is the analog of a body in equilibrium whose parts are not, and of
a neutral body containing free charges of both signs.
\begin{rem}[The free current as a dipole distribution]
\label{rem:Mf} Let $\dee{\mathfrak{J}_{f}}=0$ on an open set $\mathcal{U}$
that contains $\mathcal{R}$ and is diffeomorphic to a ball, for example
a ball in $\reals^{n}$ or all of $\reals^{n}$. For $p=0$, this
holds for every $\mathfrak{J}_{f}$, since $\mathfrak{J}_{f}$ is
then an $n$-form. The degree of $\mathfrak{J}_{f}$ is $n-p\geq1$,
so, by the Poincaré lemma, there is an $(n-p-1)$-form $\mathfrak{g}_{f}$
on $\mathcal{U}$ such that 
\begin{equation}
\mathfrak{J}_{f}=-\dee{\mathfrak{g}_{f}}.\label{eq:gf}
\end{equation}
It is unique up to the addition of a closed form. That is, $\mathfrak{J}_{f}$
is the bound current of the dipole form $\mathfrak{g}_{f}$, or $\mathfrak{g}_{f}$
is a potential for $\mathfrak{J}_{f}$. For $n=3$, $p=0$, (\ref{eq:gf})
becomes $r=-\nabla\cdot\v P_{f}$, and, for $n=3$, $p=1$, it becomes
$\v j_{f}=\nabla\cross\v M_{f}$ {[}GMS26b, Remark~5.3{]}. For $n=4$,
$p=1$, it becomes $\rho_{f}=-\nabla\cdot\v P_{f}$ and $\v j_{f}=\bdry_{t}\v P_{f}+\nabla\cross\v M_{f}$
(Section~\ref{sec:st}). By (\ref{eq:dgalpha}) with $\mathfrak{g}_{f}$
in place of $\mathfrak{g}$ and Stokes' theorem, 
\begin{equation}
\int_{\mathcal{R}}\mathfrak{J}_{f}\wedge\ga=-\int_{\mathcal{R}}\dee{\mathfrak{g}_{f}}\wedge\ga=\eps\int_{\mathcal{R}}\mathfrak{g}_{f}\wedge\mathfrak{f}-\int_{\bdry\mathcal{R}}\mathfrak{g}_{f}\wedge\ga.\label{eq:freeasg}
\end{equation}
Thus, by (\ref{eq:pvpnew}), the free current has the energy of the
dipole distribution $\mathfrak{g}_{f}$, stripped of the energy of
its surface current $\mathfrak{g}_{f}|_{\bdry\mathcal{R}}$. Writing
$\mathfrak{J}_{f}=-\dee{\mathfrak{g}_{f}}$ does not turn the free
current into dipoles. The form $\mathfrak{g}_{f}$ is not determined
by the current system, and the free current is the part of the current
that exposes no current on cuts. The first term on the right of (\ref{eq:freeasg})
is gauge invariant. Hence, the whole gauge dependence of the energy
of the free current sits in the boundary term, the energy of the missing
surface current. For $p\geq1$, it vanishes if the free current does
not cross $\bdry\mathcal{R}$, in agreement with (\ref{eq:gaugebody}).
\end{rem}

\begin{rem}[Conservation of the free current]
\label{rem:charge} Let $p\geq1$. Gauge invariance is closely related
to the conservation of the current, and the relation becomes precise
if we restrict the gauge forms. Let $S$ be any variational stress,
weakly gauge invariant or not, and let $b$ be its body force, identified
with an $(n-p)$-form $\mathfrak{J}$ by Lemma~\ref{lem:wedge},
so that, by (\ref{eq:var}), 
\[
U_{\mathcal{R}}(\ga)=\int_{\mathcal{R}}\mathfrak{J}\wedge\ga+\int_{\bdry\mathcal{R}}s\cdot\ga.
\]
Let $\gl$ be a $(p-1)$-form that vanishes in a neighborhood of $\bdry\mathcal{R}$.
Then the boundary term drops out for $\ga=\dee{\gl}$, and, as in
(\ref{eq:gaugebody}), 
\begin{equation}
U_{\mathcal{R}}(\dee{\gl})=\int_{\mathcal{R}}\mathfrak{J}\wedge\dee{\gl}=\eps\int_{\mathcal{R}}\dee{\mathfrak{J}}\wedge\gl.\label{eq:gaugecharge}
\end{equation}
By Lemma~\ref{lem:wedge}, with $k=p-1$, it follows that the energy
of every region is invariant under all gauge transformations that
vanish near its boundary if and only if 
\begin{equation}
\dee{\mathfrak{J}}=0,\label{eq:conservation}
\end{equation}
i.e., if and only if the total current is conserved. We call this
property interior gauge invariance. For a weakly gauge-invariant current
system, $\dee{\mathfrak{J}}=\dee{\mathfrak{J}_{f}}-\dee{\dee{\mathfrak{g}}}=\dee{\mathfrak{J}_{f}}$.
Thus, the bound current always conserves itself, and interior gauge
invariance reduces to the conservation of the free current, $\dee{\mathfrak{J}_{f}}=0$.
The meaning of (\ref{eq:conservation}) is that the current carried
through the boundary of every $(n-p+1)$-dimensional region vanishes,
by Stokes' theorem. For $n=4$, $p=1$, it is the conservation of
charge, $\bdry_{t}\rho+\nabla\cdot\v j=0$ (Section~\ref{sec:st}).
Hence, (\ref{eq:gaugecharge}) is the familiar correspondence between
gauge invariance and conservation of charge, here derived for the
energy of a region of spacetime and without a variational principle
for the field. For $n=3$, $p=1$, it is the condition $\nabla\cdot\v j=0$
of magnetostatics {[}GMS26b, Remark~5.4{]}. Conservation of the free
current is also exactly the hypothesis of Remark~\ref{rem:Mf}. Gauge
invariance on every region (Definition~\ref{def:gauge}) is much
stronger. By (\ref{eq:gaugebody}), it also rules out free current
that crosses the boundary of any region, and, by Theorem~\ref{thm:main},
it rules out free current altogether.

For $p=0$, there is nothing to restrict, since a constant that vanishes
near $\bdry\mathcal{R}$ vanishes on the connected components of $\mathcal{R}$
that meet $\bdry\mathcal{R}$. Accordingly, $\dee{\mathfrak{J}}=0$
holds identically, $\mathfrak{J}$ being an $n$-form.
\end{rem}

\subsection{Three levels of gauge invariance}

\label{sec:levels}

We now have three conditions on the energy of a current system, each
stronger than the previous one.
\begin{enumerate}
\item[(1)] Weak gauge invariance (Assumption~\ref{ass:weak}) holds if and
only if the variational stress has the form (\ref{eq:SJg}), i.e.,
if and only if the traction stress is represented by a dipole form
$\mathfrak{g}$ (Theorem~\ref{thm:decomp}). This is the analog of
the symmetry of the stress, and it holds automatically for $p=0$.
\item[(2)] For a weakly gauge-invariant current system and $p\geq1$, interior
gauge invariance, i.e., invariance of the energy of every region under
all gauge transformations that vanish near its boundary, holds if
and only if, in addition, $\dee{\mathfrak{J}_{f}}=0$ (Remark~\ref{rem:charge}).
Physically, the free current is conserved. Equivalently, by Remark~\ref{rem:Mf},
on an open set diffeomorphic to a ball, there is a free current potential
$\mathfrak{g}_{f}$ such that $\mathfrak{J}_{f}=-\dee{\mathfrak{g}_{f}}$.
Equivalently also, there is a total current potential $\mathfrak{g}_{T}:=\mathfrak{g}+\mathfrak{g}_{f}$
such that the total current satisfies 
\begin{equation}
\mathfrak{J}=\mathfrak{J}_{f}+\mathfrak{J}_{b}=-\dee{\mathfrak{g}_{T}}.\label{eq:gT}
\end{equation}
Indeed, by (\ref{eq:Jb}), $\mathfrak{J}=-\dee{\mathfrak{g}_{f}}-\dee{\mathfrak{g}}=-\dee{(\mathfrak{g}+\mathfrak{g}_{f})}$.
For $p=0$, this level coincides with the first.
\item[(3)] For a weakly gauge-invariant current system, gauge invariance (Definition~\ref{def:gauge})
holds if and only if, in addition, $\mathfrak{J}_{f}=0$ (Theorem~\ref{thm:main}).
Physically, there is no free current. This is the analog of the balance
of forces.
\end{enumerate}
For magnetostatics, these are the three levels of {[}GMS26b, Sect.~5.3{]},
and, for electromagnetism in spacetime, the second level is the conservation
of free charge. In Section~\ref{sec:mech}, we will work at the second
level.

\textbf{The total current potential and the excitation.} Equation
(\ref{eq:gT}) has the form of the inhomogeneous premetric Maxwell
equation {[}HO03{]}. Setting $\mathcal{H}_{T}:=-\mathfrak{g}_{T}$,
it reads $\dee{\mathcal{H}_{T}}=\mathfrak{J}$, with the total current
on the right. Equivalently, the form $\mathcal{H}:=\mathcal{H}_{T}+\mathfrak{g}=-\mathfrak{g}_{f}$
satisfies $\dee{\mathcal{H}}=\mathfrak{J}_{f}$, with the free current
on the right, and the dipole form adds to the excitation. This is
the point of view of {[}Seg16, Sect.~7(b){]}, where the Maxwell form
$\mathfrak{g}$ of that paper plays the role of $\mathcal{H}_{T}$,
up to sign conventions. Here, no constitutive relation is used, and
the potentials $\mathfrak{g}_{f}$ and $\mathfrak{g}_{T}$ are determined
by the currents only up to closed forms. A constitutive relation between
$\mathcal{H}_{T}$ and $\mathfrak{f}$ selects one of them and turns
(\ref{eq:gT}) into a field equation. For $n=4$, $p=1$, with the
aether relations, $\mathcal{H}_{T}$ is the vacuum excitation, represented
by $\eps_{0}\v E$ and $\v B/\mu_{0}$, and $\mathcal{H}$ is the
excitation in matter, represented by $\v D=\eps_{0}\v E+\v P$ and
$\v H=\v B/\mu_{0}-\v M$. The equations $\dee{\mathcal{H}_{T}}=\mathfrak{J}$
and $\dee{\mathcal{H}}=\mathfrak{J}_{f}$ are then the two familiar
forms of the inhomogeneous Maxwell equations.

\section{Mechanical forces on matter}

\label{sec:mech}

The forces exerted by the field on the matter in which the currents
are embedded are obtained by a second application of the stress theory
of Section~\ref{sec:prelim}, now with the material velocity as the
generalized velocity, i.e., with $W=T\mathcal{X}$.

\subsection{Embedded currents}

Let $\mathcal{R}$ be a region and $\dis\colon\reals\times\mathcal{R}\to\mathcal{X}$
a virtual motion with parameter $\tau$, such that $\dis_{0}=\mathrm{Id}$
and the configurations $\dis_{\tau}:=\dis(\tau,\cdot)$ are orientation-preserving
embeddings. The virtual velocity is the vector field 
\[
w:=\bdry_{\tau}\dis_{\tau}\big|_{\tau=0}
\]
on $\mathcal{R}$. For $n=4$, $\mathcal{R}$ is a region of spacetime,
$\dis_{\tau}$ deforms the family of world tubes of the material points,
and $w$ has a time component.

We say that the currents are embedded in matter if both $\mathfrak{J}_{f}$
and $\mathfrak{g}$ are convected with the material as differential
forms, while the potential $\ga$, and hence the field $\mathfrak{f}$,
is held fixed. That is, in the configuration $\dis_{\tau}$, the currents
are given on $\dis_{\tau}(\mathcal{R})$ by 
\begin{equation}
\mathfrak{J}_{f\,\tau}:=(\dis^{-1}_{\tau})^{*}\mathfrak{J}_{f},\qquad\mathfrak{g}_{\tau}:=(\dis^{-1}_{\tau})^{*}\mathfrak{g}.\label{eq:convect}
\end{equation}
Since the integral of a form is invariant under orientation-preserving
diffeomorphisms, $\int_{\dis_{\tau}(\mathcal{C})}\mathfrak{J}_{f\,\tau}=\int_{\mathcal{C}}\mathfrak{J}_{f}$
for every $(n-p)$-dimensional smooth material surface $\mathcal{C}$.
Thus, the free current through every material surface is conserved
in the motion. For $p=0$, this is the conservation of the free charge
of every material subbody. Since pullbacks commute with the exterior
derivative, $\dee{\mathfrak{J}_{f\,\tau}}=(\dis^{-1}_{\tau})^{*}\dee{\mathfrak{J}_{f}}$,
so a conserved free current remains conserved in the motion. In $\rthree$,
(\ref{eq:convect}) reproduces the embedding of {[}GMS26a, GMS26b{]}.
The free charge is convected as a 3-form, the polarization as a 2-form,
by the Piola transformation, the free current as a 2-form, again by
the Piola transformation, and the magnetization as a 1-form, by $\v M_{\tau}=\v F^{-\mathsf{T}}_{\tau}\v M$
(Section~\ref{sec:R3}).

What is embedded is the flow of the free current, not the individual
carriers. In a wire attached to a body, the charge carriers move along
the wire, relative to the material. What the embedding assumption
says is that the streamlines of the free current are carried with
the material, and that the current through every material cross-section
is not affected by the motion. The assumption rules out conduction
in a deforming conductor, where the current redistributes itself relative
to the material as the body deforms. As in {[}GMS26b{]}, we do not
consider the electromotive forces induced by the motion, i.e., the
currents are held fixed by whatever maintains them.

\textbf{Conservation of the free current.} From now on, we assume
that the free current is conserved, $\dee{\mathfrak{J}_{f}}=0$, i.e.,
we work at the second level of Section~\ref{sec:levels}. For $n=4$,
$p=1$, it is the conservation of free charge. We will point out where
the assumption is used.

\subsection{The force functional}

The energy (\ref{eq:U}) of the convected currents at the parameter
value $\tau$ is 
\begin{equation}
U_{\tau}=\int_{\dis_{\tau}(\mathcal{R})}\big(\mathfrak{J}_{f\,\tau}\wedge\ga+\eps\,\mathfrak{g}_{\tau}\wedge\mathfrak{f}\big)=\int_{\mathcal{R}}\mathfrak{J}_{f}\wedge\dis^{*}_{\tau}\ga+\eps\int_{\mathcal{R}}\mathfrak{g}\wedge\dis^{*}_{\tau}\mathfrak{f}.\label{eq:Ut}
\end{equation}
To get the second expression, we pulled the integrand back by the
orientation-preserving embedding $\dis_{\tau}$, and we used $\dis^{*}_{\tau}(\dis^{-1}_{\tau})^{*}=\mathrm{Id}$
and the fact that pullbacks commute with the wedge product. Thus,
convecting the currents in a fixed potential is equivalent to keeping
the currents fixed in the pulled-back potential. The derivative of
the pullback at $\tau=0$ is the Lie derivative, 
\[
\frac{\dee{}}{\dee{\tau}}\dis^{*}_{\tau}\ga\Big|_{\tau=0}=\mathcal{L}_{w}\ga,\qquad\frac{\dee{}}{\dee{\tau}}\dis^{*}_{\tau}\mathfrak{f}\Big|_{\tau=0}=\mathcal{L}_{w}\mathfrak{f}.
\]
By Cartan's formula 
\[
\mathcal{L}_{w}\ggm=w\mathbin{\lrcorner}\dee{\ggm}+\dee{(w\mathbin{\lrcorner}\ggm)},
\]
for any form $\ggm$. Together with $\dee{\ga}=\mathfrak{f}$ and
$\dee{\mathfrak{f}}=0$, we have, 
\begin{equation}
\mathcal{L}_{w}\ga=w\mathbin{\lrcorner}\mathfrak{f}+\dee{(w\mathbin{\lrcorner}\ga)},\qquad\mathcal{L}_{w}\mathfrak{f}=\dee{(w\mathbin{\lrcorner}\mathfrak{f})}.\label{eq:cartan}
\end{equation}
In a chart, the exterior derivative, the interior product and the
Lie derivative are given by (\ref{eq:dcoord})\textendash (\ref{eq:liecoord}).
As in {[}GMS26a, GMS26b{]}, we define the force functional as 
\begin{equation}
F_{\mathcal{R}}(w):=-\frac{\dee{}}{\dee{\tau}}U_{\tau}\Big|_{\tau=0}=-U_{\mathcal{R}}(\mathcal{L}_{w}\ga).\label{eq:F}
\end{equation}
Substituting (\ref{eq:cartan}), 
\begin{equation}
\begin{split}F_{\mathcal{R}}(w) & =\int_{\mathcal{R}}S_{M}\cdot j^{1}w,\\
S_{M}\cdot j^{1}w & :=-\mathfrak{J}_{f}\wedge(w\mathbin{\lrcorner}\mathfrak{f})-\mathfrak{J}_{f}\wedge\dee{(w\mathbin{\lrcorner}\ga)}-\eps\,\mathfrak{g}\wedge\dee{(w\mathbin{\lrcorner}\mathfrak{f})}.
\end{split}
\label{eq:Sm}
\end{equation}
Since $\ga$ and $\mathfrak{f}$ are fixed, the value of $w\mathbin{\lrcorner}\mathfrak{f}$
at a point depends on $w$ at that point, and the values of $\dee{(w\mathbin{\lrcorner}\ga)}$
and $\dee{(w\mathbin{\lrcorner}\mathfrak{f})}$ depend on $w$ and
its first derivatives. Hence, $S_{M}$ is a variational stress, a
section of $(J^{1}T\mathcal{X})^{\star}$, and $\{F_{\mathcal{R}}\}$
is a force system in the sense of Section~\ref{sec:prelim}. Its
components in a chart are given in Appendix~\ref{app:coord}. The
force on the currents in $\mathcal{R}$ is the continuous superposition
of the forces on the convected free current and on the convected dipoles.

It is convenient to separate the term that involves $\ga$ itself
from those that involve only $\mathfrak{f}$. We decompose 
\begin{equation}
\begin{split}F_{\mathcal{R}} & =F^{m}_{\mathcal{R}}+F^{g}_{\mathcal{R}},\text{ where,}\\
F^{m}_{\mathcal{R}}(w) & :=-\int_{\mathcal{R}}\mathfrak{J}_{f}\wedge(w\mathbin{\lrcorner}\mathfrak{f})-\eps\int_{\mathcal{R}}\mathfrak{g}\wedge\dee{(w\mathbin{\lrcorner}\mathfrak{f})},\\
F^{g}_{\mathcal{R}}(w) & :=-\int_{\mathcal{R}}\mathfrak{J}_{f}\wedge\dee{(w\mathbin{\lrcorner}\ga)}.
\end{split}
\label{eq:Fsplit}
\end{equation}

\subsection{Gauge dependence of the forces}
\begin{prop}
\label{prop:forcegauge} Let $\mathcal{R}$ be a region. Under a gauge
transformation $\ga\mapsto\ga+\gb$, for a closed form $\gb$, the
force functional changes by 
\begin{equation}
\gd F_{\mathcal{R}}(w)=-\int_{\mathcal{R}}\mathfrak{J}_{f}\wedge\dee{(w\mathbin{\lrcorner}\gb)},\label{eq:dF}
\end{equation}
which, by the conservation of the free current, may be written as
\begin{equation}
\gd F_{\mathcal{R}}(w)=\eps\int_{\bdry\mathcal{R}}\mathfrak{J}_{f}\wedge(w\mathbin{\lrcorner}\gb).\label{eq:dFb}
\end{equation}
Consequently:
\begin{enumerate}
\item[(a)] $F^{m}_{\mathcal{R}}$, which contains the force on the dipoles,
is gauge invariant.
\item[(b)] For $p=0$, the force functional is gauge invariant for every current
system.
\item[(c)] For $p\geq1$, $F_{\mathcal{R}}(w)$ is gauge invariant for every
$w$ if and only if the free current does not cross $\bdry\mathcal{R}$,
and it is gauge invariant for every $w$ that vanishes on $\bdry\mathcal{R}$.
\end{enumerate}
\end{prop}

\begin{proof}
The field $\mathfrak{f}$ is gauge invariant, so the only term of
(\ref{eq:Fsplit}) that changes is $F^{g}_{\mathcal{R}}$, and (\ref{eq:dF})
follows. Equation (\ref{eq:dFb}) is (\ref{eq:gaugebody}) with $\gl$
replaced by the $(p-1)$-form $w\mathbin{\lrcorner}\gb$ and $\dee{\mathfrak{J}_{f}}=0$.
Statement (a) is immediate, and (b) holds since the interior product
of a vector field with a 0-form vanishes. For (c), if the free current
does not cross $\bdry\mathcal{R}$, the integrand of (\ref{eq:dFb})
vanishes, and so does $\gd F_{\mathcal{R}}(w)$. If $w$ vanishes
on $\bdry\mathcal{R}$, then so does $w\mathbin{\lrcorner}\gb$, and
again $\gd F_{\mathcal{R}}(w)=0$. Conversely, assume that the restriction
of $\mathfrak{J}_{f}$ to $\bdry\mathcal{R}$ does not vanish at some
point. By Lemma~\ref{lem:wedge}, applied on $\bdry\mathcal{R}$,
there is a $(p-1)$-form on $\bdry\mathcal{R}$ whose wedge product
with the restriction of $\mathfrak{J}_{f}$ does not vanish at that
point, and, as in the explanation following (\ref{eq:gaugebody}),
it may be localized so that the integral in (\ref{eq:dFb}) does not
vanish. Hence, to prove that the force is not gauge invariant, it
suffices to show that the $(p-1)$-form $w\mathbin{\lrcorner}\gb$
can be made equal to any given $(p-1)$-form near a point of $\bdry\mathcal{R}$,
by the choice of the closed form $\gb$ and the vector field $w$.
In a chart, let $\psi$ be a $(p-1)$-form with constant coefficients
that does not contain $\dee{x^{m}}$, let $h$ be a function supported
in the chart, and set $w:=h\,\bdry_{m}$ and $\gb:=\dee{(\phi\,x^{m}\psi)}$,
where $\phi$ is a cutoff function supported in the chart and equal
to $1$ on a neighborhood of the support of $h$. Then $\gb$ is exact,
hence closed, on $\mathcal{X}$, it equals $\dee{x^{m}}\wedge\psi$
where $h\neq0$, and $w\mathbin{\lrcorner}\gb=h\psi$. Every $(p-1)$-form
is a sum of such terms, and the conclusion follows from (\ref{eq:dFb})
and Lemma~\ref{lem:wedge}, applied on $\bdry\mathcal{R}$.
\end{proof}

This is the main difference between electrostatics and the cases $p\geq1$.
For $p=0$, a gauge transformation changes the energy of the convected
free charge by $c\int_{\mathcal{R}}\mathfrak{J}_{f}$, a constant
that does not change in the motion. As a result, the force is gauge
invariant for every charge system {[}GMS26a{]}. For $p\geq1$, by
(\ref{eq:Ut}), it changes the energy by $\int_{\mathcal{R}}\mathfrak{J}_{f}\wedge\dis^{*}_{\tau}\gb$,
which, in general, changes in the motion. If the free current does
not cross $\bdry\mathcal{R}$, then, by (\ref{eq:gaugebody}) applied
to $\dis^{*}_{\tau}\gb=\dee{(\dis^{*}_{\tau}\gl)}$, it vanishes for
all $\tau$. Otherwise, the motion of the points where the free current
crosses $\bdry\mathcal{R}$ requires the rest of the circuit to follow,
and the energy of that part of the circuit, which lies outside $\mathcal{R}$,
is not included in $U_{\mathcal{R}}$. This is the origin of the dependence
on the gauge. Virtual velocities that vanish on $\bdry\mathcal{R}$
do not move these points, and, for them, the force is gauge invariant.

\textbf{Gauge invariance of the forces on all regions.} We now require
the force to be gauge invariant on every region and for every virtual
velocity. Let $p\geq1$. Then (\ref{eq:dF}) vanishes for every $\mathcal{R}$,
so that $\mathfrak{J}_{f}\wedge\dee{(w\mathbin{\lrcorner}\gb)}=0$
pointwise. With $\gb=\dee{x^{m}}\wedge\dee{x^{I'}}$, where $I'$
is a multi-index of length $p-1$ that does not contain $m$, and
$w=h\,\bdry_{m}$, extended as in the proof of Proposition~\ref{prop:forcegauge},
we get $\dee{(w\mathbin{\lrcorner}\gb)}=\dee h\wedge\dee{x^{I'}}$,
and, since $h$ is arbitrary, $\mathfrak{J}_{f}\wedge\dee{x^{k}}\wedge\dee{x^{I'}}=0$
for every $k$. Every $\dee{x^{K}}$ with $|K|=p$ is of the form
$\pm\dee{x^{k}}\wedge\dee{x^{I'}}$, and, since $p-1<n$, an index
$m\notin I'$ can always be chosen. Hence, $\mathfrak{J}_{f}\wedge\dee{x^{K}}=0$
for all $K$, and $\mathfrak{J}_{f}=0$ by Lemma~\ref{lem:wedge}.
We conclude that the forces are gauge invariant on all regions if
and only if the energy is, i.e., if and only if there is no free current.
For the forces, there is thus no intermediate condition between gauge
invariance on all regions and no gauge invariance at all, as long
as all virtual velocities are admitted. An intermediate condition,
analogous to interior gauge invariance of the energy, is obtained
by admitting only virtual velocities that vanish near the boundary
of the region, as we now show.

\textbf{The role of conservation.} The conservation of the free current
is what makes the force gauge invariant for virtual velocities that
vanish near the boundary. Without assuming it, if $w$ vanishes in
a neighborhood of $\bdry\mathcal{R}$, then so does $w\mathbin{\lrcorner}\gb$,
and (\ref{eq:dF}) and (\ref{eq:gaugebody}) give 
\[
\gd F_{\mathcal{R}}(w)=-\eps\int_{\mathcal{R}}\dee{\mathfrak{J}_{f}}\wedge(w\mathbin{\lrcorner}\gb).
\]
Without conservation, this does not vanish in general, and the force
would depend on the gauge even in the interior of $\mathcal{R}$.
With it, the only gauge dependence left is the boundary term (\ref{eq:dFb}),
due to the free current that crosses $\bdry\mathcal{R}$.

\subsection{Interpretation as a force system}

\label{sec:forcesystem}
\begin{prop}
\label{prop:mech} Let $\mathcal{R}$ be a region.
\begin{enumerate}
\item[(a)] The force system $F^{m}$ has the traction stress $\gs_{m}$ and
the body force $b_{m}$ given by 
\begin{equation}
\gs_{m}\cdot w:=-\mathfrak{g}\wedge(w\mathbin{\lrcorner}\mathfrak{f}),\qquad b_{m}\cdot w:=-\mathfrak{J}\wedge(w\mathbin{\lrcorner}\mathfrak{f}),\label{eq:smbm}
\end{equation}
where $\mathfrak{J}=\mathfrak{J}_{f}-\dee{\mathfrak{g}}$ is the total
current, and 
\begin{equation}
F^{m}_{\mathcal{R}}(w)=-\int_{\mathcal{R}}\mathfrak{J}\wedge(w\mathbin{\lrcorner}\mathfrak{f})-\int_{\bdry\mathcal{R}}\mathfrak{g}\wedge(w\mathbin{\lrcorner}\mathfrak{f}).\label{eq:Fmatter}
\end{equation}
\item[(b)] If the free current is conserved, $\dee{\mathfrak{J}_{f}}=0$, the
force system $F^{g}$ has the traction stress $\gs_{g}$ given by
\begin{equation}
\gs_{g}\cdot w:=\eps\,\mathfrak{J}_{f}\wedge(w\mathbin{\lrcorner}\ga),\label{eq:sigmag}
\end{equation}
and no body force. Thus, 
\begin{equation}
F^{g}_{\mathcal{R}}(w)=\eps\int_{\bdry\mathcal{R}}\mathfrak{J}_{f}\wedge(w\mathbin{\lrcorner}\ga).\label{eq:Fg}
\end{equation}
\end{enumerate}
In particular, if $\dee{\mathfrak{J}_{f}}=0$, the variational stress
$S_{M}$ of (\ref{eq:Sm}) decomposes in the form 
\begin{equation}
S_{M}=\mathfrak{d}(\gs_{m}+\gs_{g})+b_{m}\circ\pi^{1}_{0},\quad\text{i.e.,}\quad S_{M}\cdot j^{1}w=\dee{\big((\gs_{m}+\gs_{g})\cdot w\big)}+b_{m}\cdot w,\label{eq:Smdecomp}
\end{equation}
and 
\begin{equation}
F_{\mathcal{R}}(w)=-\int_{\mathcal{R}}\mathfrak{J}\wedge(w\mathbin{\lrcorner}\mathfrak{f})-\int_{\bdry\mathcal{R}}\mathfrak{g}\wedge(w\mathbin{\lrcorner}\mathfrak{f})+\eps\int_{\bdry\mathcal{R}}\mathfrak{J}_{f}\wedge(w\mathbin{\lrcorner}\ga).\label{eq:pvpm}
\end{equation}
\end{prop}

\begin{proof}
(a) Since $\mathfrak{g}$ is of degree $n-p-1$, the Leibniz rule
(\ref{eq:leibniz}) gives 
\[
\dee{\big(\mathfrak{g}\wedge(w\mathbin{\lrcorner}\mathfrak{f})\big)}=\dee{\mathfrak{g}}\wedge(w\mathbin{\lrcorner}\mathfrak{f})+\eps\,\mathfrak{g}\wedge\dee{(w\mathbin{\lrcorner}\mathfrak{f})}.
\]
Substituting $-\eps\,\mathfrak{g}\wedge\dee{(w\mathbin{\lrcorner}\mathfrak{f})}=-\dee{\big(\mathfrak{g}\wedge(w\mathbin{\lrcorner}\mathfrak{f})\big)}+\dee{\mathfrak{g}}\wedge(w\mathbin{\lrcorner}\mathfrak{f})$
in (\ref{eq:Fsplit}), the integrand of $F^{m}_{\mathcal{R}}$ is
\[
-\dee{\big(\mathfrak{g}\wedge(w\mathbin{\lrcorner}\mathfrak{f})\big)}-(\mathfrak{J}_{f}-\dee{\mathfrak{g}})\wedge(w\mathbin{\lrcorner}\mathfrak{f})=\mathfrak{d}(\gs_{m})\cdot j^{1}w+b_{m}\cdot w.
\]
The second term is of zeroth order. By (P1) and (P2), the traction
stress is $\gs_{m}$. By (\ref{eq:div}) and (\ref{eq:body}), the
body force is $b_{m}$. Equation (\ref{eq:Fmatter}) follows by Stokes'
theorem.

(b) As in the derivation of (\ref{eq:gaugebody}), with the $(p-1)$-form
$w\mathbin{\lrcorner}\ga$ in place of $\gl$, 
\[
\begin{split}-\mathfrak{J}_{f}\wedge\dee{(w\mathbin{\lrcorner}\ga)} & =\eps\,\dee{\big(\mathfrak{J}_{f}\wedge(w\mathbin{\lrcorner}\ga)\big)}-\eps\,\dee{\mathfrak{J}_{f}}\wedge(w\mathbin{\lrcorner}\ga)\\
 & =\mathfrak{d}(\gs_{g})\cdot j^{1}w,
\end{split}
\]
where we used $\dee{\mathfrak{J}_{f}}=0$. Hence, by (P2), the traction
stress is $\gs_{g}$. The body force vanishes, since $\mathrm{div}\,\mathfrak{d}(\gs_{g})=0$.
Equation (\ref{eq:Fg}) follows by Stokes' theorem. For $p=0$, both
sides vanish.

Adding (a) and (b) gives (\ref{eq:Smdecomp}) and (\ref{eq:pvpm}).
\end{proof}

\textbf{The gauge-invariant part.} The body force $b_{m}$ is the
Lorentz-type force exerted by the field on the total current $\mathfrak{J}$,
free plus bound, and it depends on $\ga$ only through $\mathfrak{f}$.
The traction of $F^{m}$ is the force on the surface current $\mathfrak{g}|_{\bdry\mathcal{R}}$
exposed by the dipoles on $\bdry\mathcal{R}$. The two representations
of $F^{m}_{\mathcal{R}}$, (\ref{eq:Fsplit}) by a variational stress
acting on $j^{1}w$ and (\ref{eq:Fmatter}) by a body force and a
traction, are the general counterparts of the two computations of
the force on a current loop in {[}GMS26b, Sect.~2{]}, the first through
the magnetic moment and the second through the individual current
elements.

\textbf{The gauge-dependent part.} The force system $F^{g}$ has a
traction stress, $\gs_{g}$, carried by the free current, no body
force, and its traction acts only where the free current crosses $\bdry\mathcal{R}$.
It depends on the gauge, as it should by Proposition~\ref{prop:forcegauge}.
Moreover, it is self-equilibrated in the following sense. Let a region
$\mathcal{R}$ be divided by a smooth hypersurface $\Sigma$ into
two regions $\mathcal{R}_{1}$ and $\mathcal{R}_{2}$. The orientations
that $\Sigma$ inherits as a part of $\bdry\mathcal{R}_{1}$ and as
a part of $\bdry\mathcal{R}_{2}$ are opposite. Hence, the contributions
of $\Sigma$ to $F^{g}_{\mathcal{R}_{1}}$ and $F^{g}_{\mathcal{R}_{2}}$
cancel, and 
\[
F^{g}_{\mathcal{R}_{1}}(w)+F^{g}_{\mathcal{R}_{2}}(w)=F^{g}_{\mathcal{R}}(w)
\]
for every $w$. If, in addition, the free current does not cross $\bdry\mathcal{R}$,
then $F^{g}_{\mathcal{R}}(w)=0$ for every $w$, by (\ref{eq:Fg}).
Thus, adding the traction stress of $F^{g}$ changes neither the body
force nor the total force on such a region. This is analogous to the
familiar static indeterminacy of continuum mechanics. The loads on
a body do not determine the stress in it. Here, for a given potential,
the traction stress $\gs_{g}$ is determined, but it changes under
gauge transformations, while the body force and the total force on
every region that the free current does not cross remain unchanged.
Thus, the field $\mathfrak{f}$ alone does not determine the interactions
across cuts through the free current.

\textbf{What the field determines, and what it does not.} The body
force $b_{m}$ is gauge invariant, and it is determined by the field,
whether or not the free current crosses $\bdry\mathcal{R}$. Indeed,
if $w$ vanishes in a neighborhood of $\bdry\mathcal{R}$, the boundary
terms of (\ref{eq:pvpm}) drop out, and 
\[
F_{\mathcal{R}}(w)=-\int_{\mathcal{R}}\mathfrak{J}\wedge(w\mathbin{\lrcorner}\mathfrak{f})
\]
in every gauge. Such virtual velocities can be concentrated about
any interior point of $\mathcal{R}$, so they determine the body force
everywhere. What the field does not determine is the traction on the
parts of $\bdry\mathcal{R}$ through which the free current flows.
For a given potential $\ga$, this traction is determined by the energy,
and it is given by (\ref{eq:Fg}). However, it changes under gauge
transformations, which do not change the field, and, by Proposition~\ref{prop:forcegauge},
it is gauge invariant only where it vanishes. We will see below that
it is not a force on matter.

\textbf{The gauge-dependent part as a flux of potential energy-momentum.}
The meaning of the traction stress $\gs_{g}$ of Proposition~\ref{prop:mech}
is revealed by Cartan's formula (\ref{eq:cartan}). By (\ref{eq:F}),
the free current contributes $-\int_{\mathcal{R}}\mathfrak{J}_{f}\wedge\mathcal{L}_{w}\ga$
to the force, and (\ref{eq:cartan}) splits it into the Lorentz-type
force $-\int_{\mathcal{R}}\mathfrak{J}_{f}\wedge(w\mathbin{\lrcorner}\mathfrak{f})$
and the term $F^{g}_{\mathcal{R}}(w)$, which, by (\ref{eq:Fg}),
is a boundary term.

To see the physical meaning of the split, consider a single point
charge $q$ moving along a world line $\gamma$ in spacetime, for
the electrodynamics case $n=4$, $p=1$. Its free current is concentrated
on $\gamma$, and, with the conventions of Section~\ref{sec:st},
where this case is considered, its energy in a region $\mathcal{R}$
is 
\[
U_{\mathcal{R}}(\ga)=-q\int_{\gamma\cap\mathcal{R}}\ga=q\int^{t_{2}}_{t_{1}}\big(\vph-\v u\cdot\v A\big)\dee t,
\]
where $\v u$ is the velocity of the charge and $[t_{1},t_{2}]$ is
the time interval during which it is in $\mathcal{R}$. This is minus
the time integral of the interaction term $-q\vph+q\,\v u\cdot\v A$
of the Lagrangian of a charge in an electromagnetic field, both in
the nonrelativistic case {[}GPS02, Sect.~1.5{]} and in the relativistic
one {[}Jac99, Sect.~12.1{]}. When $\gamma$ is deformed with the
virtual velocity $w=w^{0}\bdry_{t}+\v w$, (\ref{eq:F}) and (\ref{eq:cartan})
give 
\[
F_{\mathcal{R}}(w)=q\int_{\gamma\cap\mathcal{R}}\mathcal{L}_{w}\ga=q\int_{\gamma\cap\mathcal{R}}w\mathbin{\lrcorner}\mathfrak{f}+q\,\big[w\mathbin{\lrcorner}\ga\big]^{\text{exit}}_{\text{entry}},
\]
where the last term is the difference between the values of the function
$w\mathbin{\lrcorner}\ga$ at the points where $\gamma$ leaves and
enters $\mathcal{R}$. By (\ref{eq:wf4}) and (\ref{eq:wa4}) below,
the first term is 
\[
q\int^{t_{2}}_{t_{1}}\Big(\big(\v E+\v u\cross\v B\big)\cdot\v w-(\v E\cdot\v u)\,w^{0}\Big)\dee t,
\]
the virtual work of the Lorentz force, and of minus the power of the
field, along the world line. The second term is 
\[
q\,\big[\v A\cdot\v w-\vph\,w^{0}\big]^{\text{exit}}_{\text{entry}}.
\]
In Lagrangian mechanics, the variation of the action under a variation
of the path that moves its end points by $\gd\v x$ and $\gd t$ contains
the end-point terms $\v p\cdot\gd\v x-H\gd t$, where $\v p$ is the
canonical momentum and $H$ the canonical energy (see, e.g., {[}GPS02,
Ch.~10{]}). For a charge in an electromagnetic field, $\v p=\v p_{\mathrm{kin}}+q\v A$
and $H=H_{\mathrm{kin}}+q\vph$, where $\v p_{\mathrm{kin}}=\gamma m\v u$
and $H_{\mathrm{kin}}=\gamma mc^{2}$ are the mechanical momentum
and energy of the charge, with mass $m$, velocity $\v u$, the speed
of light $c$, and $\gamma=(1-|\v u|^{2}/c^{2})^{-1/2}$ {[}Jac99,
Sect.~12.1{]}. Thus, $q\v A$ and $q\vph$ are the potential momentum
and the potential energy of the charge, and the end-point term above
is the covector $q\ga$, with components $(-q\vph,q\v A)$, paired
with $(w^{0},\v w)$. This is the pairing $\v p\cdot\gd\v x-H\gd t$,
in which the potential energy $q\vph$ enters with a minus sign. Thus,
the end-point term is the potential energy-momentum of the charge,
paired with the virtual displacement, as it leaves $\mathcal{R}$,
minus its value as the charge enters $\mathcal{R}$. For the free
current of a continuum of carriers, the same holds by superposition,
and the end-point terms add up to the boundary integral (\ref{eq:Fg}).

We conclude that the force derived from the energy is the Lorentz-type
force plus the net rate at which potential energy-momentum leaves
$\mathcal{R}$ with the free current. This identification is an interpretation.
Proposition~\ref{prop:mech} establishes that $F^{g}$ is a gauge-dependent
boundary term with no body force, and its reading as a flux of potential
energy-momentum rests, in addition, on the point-charge argument above
and on the embedding assumption (\ref{eq:convect}). With this interpretation,
the traction stress $\gs_{g}$ is the flux of potential energy-momentum
carried by the free current. It is not a mechanical force exerted
on the matter. The carriers exchange momentum and energy with the
material. When the inertia of the carriers is negligible, as for the
conduction electrons in a metal, the material exerts on them a force
that balances the Lorentz force, and the carriers pass on to the material
the energy that they receive from the field. By reaction, the mechanical
force on the material is the Lorentz force, and its time component
accounts for the energy delivered to the material, for example, as
Joule heat. The flux of potential energy-momentum depends on the gauge,
because the potential energy-momentum does. Only its circulation around
closed loops, the flux of $\mathfrak{f}$, has a gauge-invariant meaning.
This is also why $F^{g}_{\mathcal{R}}$ vanishes when the free current
does not cross $\bdry\mathcal{R}$, since what flows out of a region
flows into the neighboring one. For magnetostatics, $n=3$, $p=1$,
this is the interpretation of {[}GMS26b, Sect.~6.3 and App.~A{]}.
For electrostatics, $p=0$, the interior product $w\mathbin{\lrcorner}\ga$
vanishes, and so does the flux.

\textbf{The force on matter.} We conclude that the force exerted by
the field on the matter in a region $\mathcal{R}$ is the gauge-invariant
part $F^{m}_{\mathcal{R}}$ of (\ref{eq:Fmatter}), i.e., the Lorentz-type
force on the total current and on the surface current exposed by the
dipoles. It is represented by a variational stress, it is gauge invariant,
and it is defined for every region, whether or not the free current
crosses its boundary. It agrees with the force derived from the energy
for every region that the free current does not cross, and, for every
region, for virtual velocities that vanish near its boundary.
\begin{rem}[What is invariant and what is not]
\label{rem:split} On a general manifold, $T\mathcal{X}$ carries
no natural connection, so the splitting of $S_{M}\cdot j^{1}w$, or
of the variational stress of $F^{m}$, into a part acting on the value
of $w$ and a part acting on its derivatives is not invariant {[}Seg23,
Sect.~10.8{]}. With a connection, as in {[}Seg23, Sect.~10.9{]},
the variational stress splits into a body-force-like component and
a stress acting on the covariant derivative of $w$. In $\rthree$,
with the flat connection, these are the dipole force densities and
the asymmetric active stresses of {[}GMS26a, GMS26b{]} (Section~\ref{sec:R3}).
The invariant objects are $S_{M}$ itself, its traction stress, and
its body force.
\end{rem}

\section{Electrostatics and magnetostatics in $\protect\rthree$}

\label{sec:R3}

In this section, $\mathcal{X}=\rthree$, and we show how the results
of {[}GMS26a{]} and {[}GMS26b{]} are recovered. We write forms in
terms of vector fields, using the Euclidean structure of $\rthree$.
For a vector field $\v v$, $\v v^{\flat}:=v_{i}\,\dee{x^{i}}$ denotes
the corresponding 1-form and $\v v\mathbin{\lrcorner}\dV$ the corresponding
2-form, where $\dV=\dee{x^{1}}\wedge\dee{x^{2}}\wedge\dee{x^{3}}$.
We use the identities 
\begin{align}
(\v a\mathbin{\lrcorner}\dV)\wedge\v b^{\flat} & =(\v a\cdot\v b)\dV,\label{eq:R3id}\\
\v a^{\flat}\wedge\v b^{\flat} & =(\v a\cross\v b)\mathbin{\lrcorner}\dV,\label{eq:R3idb}\\
\v v\mathbin{\lrcorner}(\v u\mathbin{\lrcorner}\dV) & =(\v u\cross\v v)^{\flat},\label{eq:R3idc}\\
\v v\mathbin{\lrcorner}\v a^{\flat} & =\v a\cdot\v v,\label{eq:R3idd}\\
\dee f & =(\nabla f)^{\flat},\label{eq:R3ide}\\
\dee{\v a^{\flat}} & =(\nabla\cross\v a)\mathbin{\lrcorner}\dV,\label{eq:R3idf}\\
\dee{(\v a\mathbin{\lrcorner}\dV)} & =(\nabla\cdot\v a)\dV,\label{eq:R3idg}
\end{align}
for vector fields $\v a$, $\v b$, $\v u$, $\v v$ and a function
$f$. In addition, on the boundary of a region with outward unit normal
$\v{\nor}$, the 2-form $\v a\mathbin{\lrcorner}\dV$ restricts to
$(\v a\cdot\v{\nor})\dA$. These identities are the standard dictionary
between forms and vector analysis in $\rthree$ (see, e.g., {[}Fra12,
Sects.~2.9\textendash 2.10{]}), and we derive them in Appendix~\ref{app:vec}.
In this section, the virtual velocity is a vector field on $\rthree$,
and, as in {[}GMS26a, GMS26b{]}, we denote it by $\v w$.

\subsection{Electrostatics, $p=0$}

Here $\eps=(-1)^{3-0-1}=1$, the potential is the function $\vph$,
the field $\mathfrak{f}=\dee{\vph}$ is represented by the electric
field $\v E=-\nabla\vph$. By Theorem~\ref{thm:decomp}, every variational
stress is weakly gauge invariant. We write the free current, a 3-form,
and the dipole form, a 2-form, as 
\[
\mathfrak{J}_{f}=r\dV,\qquad\mathfrak{g}=\v P\mathbin{\lrcorner}\dV,
\]
where $r$ is the free charge density and $\v P$ is the polarization.
By (\ref{eq:U}), (\ref{eq:R3id}) and (\ref{eq:R3ide}), the energy
assumes the form 
\begin{equation}
U_{\mathcal{R}}(\vph)=\int_{\mathcal{R}}\big(r\vph+\v P\cdot\nabla\vph\big)\dV=\int_{\mathcal{R}}r\vph\dV-\int_{\mathcal{R}}\v P\cdot\v E\dV,
\end{equation}
which is the energy of {[}GMS26a{]}. The bound current is $\mathfrak{J}_{b}=-\dee{\mathfrak{g}}=-(\nabla\cdot\v P)\dV$,
so that the total charge density is $\rho=r-\nabla\cdot\v P$. The
analog of the traction stress $\vph\mapsto\vph\,\v P\mathbin{\lrcorner}\dV$
restricts to $\vph\,(\v P\cdot\v{\nor})\dA$ on $\bdry\mathcal{R}$,
the energy of the bound surface charge $\v P\cdot\v{\nor}$. The first
two levels of gauge invariance hold automatically, and the third,
gauge invariance, holds if and only if $r=0$. By Remark~\ref{rem:Mf},
on a set which is diffeomorphic to a ball, $r=-\nabla\cdot\v P_{f}$
for a vector field $\v P_{f}$.

The convection (\ref{eq:convect}) of the 3-form $r\dV$ conserves
the free charge of every material subbody, and that of the 2-form
$\v P\mathbin{\lrcorner}\dV$ is the Piola transformation $\v P_{\tau}=J^{-1}_{\tau}\v F_{\tau}\v P$,
as in {[}GMS26a{]}. For the forces, $\v w\mathbin{\lrcorner}\mathfrak{f}=\v w\cdot\nabla\vph=-\v E\cdot\v w$
and $\v w\mathbin{\lrcorner}\ga=0$, so that (\ref{eq:Sm}) gives
\begin{equation}
\begin{split}S_{M}\cdot j^{1}\v w & =\big(r\,\v E\cdot\v w+\v P\cdot\nabla(\v E\cdot\v w)\big)\dV\\
 & =\Big(\big(r\v E+(\v P\cdot\nabla)\v E\big)\cdot\v w+(\v E\otimes\v P):\nabla\v w\Big)\dV,
\end{split}
\end{equation}
where we used $P_{i}(E_{k}w_{k})_{,i}=P_{i}E_{k,i}w_{k}+E_{k}P_{i}w_{k,i}$.
The terms are the force on the free charge, the dipole force density,
and the asymmetric active stress $\v E\otimes\v P$ of {[}GMS26a{]}.
By (\ref{eq:smbm}), the body force is $\rho\,\v E\cdot\v w\dV$,
the Coulomb force on the total charge, and the traction is $(\v P\cdot\v{\nor})\v E\cdot\v w\dA$,
the force on the bound surface charge. By Proposition~\ref{prop:forcegauge}(b),
the forces are gauge invariant for every charge system.

\subsection{Magnetostatics, $p=1$}

Here $\eps=(-1)^{3-1-1}=-1$, the potential is a $1$-form $\ga=\v A^{\flat}$,
where $\v A$ is the vector potential, and $\mathfrak{f}=\dee{\ga}=\v B\mathbin{\lrcorner}\dV$,
where $\v B=\nabla\cross\v A$ is the magnetic flux field. We write
the free current, a 2-form, and the dipole form, a 1-form, as 
\[
\mathfrak{J}_{f}=-\v j_{f}\mathbin{\lrcorner}\dV,\qquad\mathfrak{g}=\v M^{\flat},
\]
where $\v M$ is the magnetization, so that, by (\ref{eq:SJg}), (\ref{eq:R3id})
and (\ref{eq:R3idf}), 
\begin{equation}
S\cdot j^{1}\ga=\mathfrak{J}_{f}\wedge\ga-\mathfrak{g}\wedge\dee{\ga}=-\big(\v j_{f}\cdot\v A+\v M\cdot\v B\big)\dV.
\end{equation}
The minus sign in the definition of $\mathfrak{J}_{f}$ expresses
the fact that a current in the potential $\v A$ has the potential
energy $-\int\v j\cdot\v A\dV$, the continuum form of $-\v m\cdot\v B$
for a current loop {[}GMS26b, Sect.~2{]}. Writing the representative
of a general variational stress as $(A_{i},A_{i,j})\mapsto(-r_{i}A_{i}-\Sigma_{ij}A_{i,j})\dV$,
weak gauge invariance is equivalent to the skew-symmetry of $\v{\Sigma}$,
which is then represented by $\v M$ {[}GMS26b, Prop.~4.2{]}. By
Table~\ref{tab:rank}, this is the case $p=1$ of Corollary~\ref{cor:form}.

The bound current is $\mathfrak{J}_{b}=-\dee{\mathfrak{g}}=-(\nabla\cross\v M)\mathbin{\lrcorner}\dV$,
so that the total current is $\mathfrak{J}=-(\v j_{f}+\nabla\cross\v M)\mathbin{\lrcorner}\dV$,
minus the 2-form of the total current $\v j=\v j_{f}+\nabla\cross\v M$.
The traction stress is $\ga\mapsto\v M^{\flat}\wedge\v A^{\flat}=(\v M\cross\v A)\mathbin{\lrcorner}\dV$,
and it restricts on $\bdry\mathcal{R}$ to $(\v M\cross\v A)\cdot\v{\nor}\dA=-(\v M\cross\v{\nor})\cdot\v A\dA$,
the energy of the bound surface current $\v M\cross\v{\nor}$. Since
$\dee{\mathfrak{J}_{f}}=-(\nabla\cdot\v j_{f})\dV$, the three levels
of gauge invariance are the skew-symmetry of $\v{\Sigma}$, the condition
$\nabla\cdot\v j_{f}=0$, and the condition $\v j_{f}=0$, as in {[}GMS26b,
Sect.~5.3{]}. By Remark~\ref{rem:Mf}, a divergence-free $\v j_{f}$
on a ball, or on $\rthree$, is of the form $\nabla\cross\v M_{f}$.

The convection (\ref{eq:convect}) of the 2-form $\v j_{f}\mathbin{\lrcorner}\dV$
is the Piola transformation $\v j_{f\,\tau}=J^{-1}_{\tau}\v F_{\tau}\v j_{f}$,
and that of the 1-form $\v M^{\flat}$ is $\v M_{\tau}=\v F^{-\mathsf{T}}_{\tau}\v M$,
as in {[}GMS26b, Sect.~6.1{]}. For the forces, (\ref{eq:R3idc})
and (\ref{eq:R3idd}) give $\v w\mathbin{\lrcorner}\mathfrak{f}=(\v B\cross\v w)^{\flat}$
and $\v w\mathbin{\lrcorner}\ga=\v A\cdot\v w$. Hence, (\ref{eq:Sm})
gives 
\begin{equation}
S_{M}\cdot j^{1}\v w=\Big((\v j_{f}\cross\v B)\cdot\v w+\v j_{f}\cdot\nabla(\v A\cdot\v w)+\v M\cdot\nabla\cross(\v B\cross\v w)\Big)\dV.\label{eq:7-11}
\end{equation}
Since $\nabla\cdot\v B=0$, $\nabla\cross(\v B\cross\v w)=\v B(\nabla\cdot\v w)+(\v w\cdot\nabla)\v B-(\v B\cdot\nabla)\v w$,
and the last term in Equation (\ref{eq:7-11}) is 
\[
\v M\cdot\nabla\cross(\v B\cross\v w)=(\nabla\v B)\trps\v M\cdot\v w+\big((\v M\cdot\v B)\v I-\v M\otimes\v B\big):\nabla\v w,
\]
i.e., the dipole force density and the asymmetric active stress, whose
skew-symmetric part is the couple density $\v M\cross\v B$ {[}GMS26b,
Sect.~6.3{]}. By Proposition~\ref{prop:mech}, the body force is
$-\mathfrak{J}\wedge(\v w\mathbin{\lrcorner}\mathfrak{f})=(\v j\cross\v B)\cdot\v w\dV$,
the Lorentz force on the total current, and the traction of $F^{m}$
is $-\mathfrak{g}\wedge(\v w\mathbin{\lrcorner}\mathfrak{f})=\big((\v M\cross\v{\nor})\cross\v B\big)\cdot\v w\dA$,
the force on the bound surface current. The traction of $F^{g}$ is
$\eps\,\mathfrak{J}_{f}\wedge(\v w\mathbin{\lrcorner}\ga)=(\v j_{f}\cdot\v{\nor})(\v A\cdot\v w)\dA$.
It is gauge dependent, it is present only where free current crosses
$\bdry\mathcal{R}$, and it is the flux of potential momentum of {[}GMS26b,
Sect.~6.3 and App.~A{]}.

\section{Electromagnetism in spacetime: $n=4$, $p=1$}

\label{sec:st}

\subsection{The setting}

Let the four-dimensional spacetime $\mathcal{X}$ be given a frame
$\reals\times\rthree$ with coordinates $(t,x^{1},x^{2},x^{3})$.
We use the orientation induced by $\dee t\wedge\dV$, where $\dV=\dee{x^{1}}\wedge\dee{x^{2}}\wedge\dee{x^{3}}$.
No metric on spacetime is used. The Euclidean structure of $\rthree$
is used, as in Section~\ref{sec:R3}. The results, when written in
terms of forms, do not depend on the frame and on the Euclidean structure
of $\rthree$. As in Section~\ref{sec:R3}, $\v v^{\flat}$ and $\v v\mathbin{\lrcorner}\dV$
denote the 1-form and the 2-form of a time-dependent vector field
$\v v$ on $\rthree$, without $\dee t$ components. Now $\eps=(-1)^{4-1-1}=1$.
The potential is the 1-form 
\begin{equation}
\ga=-\vph\dee t+\v A^{\flat},
\end{equation}
and the field is the Faraday 2-form 
\begin{equation}
\mathfrak{f}=\dee{\ga}=\v E^{\flat}\wedge\dee t+\v B\mathbin{\lrcorner}\dV,\qquad\v E=-\nabla\vph-\bdry_{t}\v A,\quad\v B=\nabla\cross\v A.\label{eq:faraday}
\end{equation}
Indeed (see the end of Appendix~\ref{app:vec}), since $\dee t\wedge\dee t=0$,
$\dee{(-\vph\dee t)}=-(\nabla\vph)^{\flat}\wedge\dee t$, and 
\[
\dee{\v A^{\flat}}=\dee t\wedge(\bdry_{t}\v A)^{\flat}+(\nabla\cross\v A)\mathbin{\lrcorner}\dV=-(\bdry_{t}\v A)^{\flat}\wedge\dee t+\v B\mathbin{\lrcorner}\dV.
\]
Collecting the terms with $\dee t$, we get $-(\nabla\vph+\bdry_{t}\v A)^{\flat}\wedge\dee t=\v E^{\flat}\wedge\dee t$.
A gauge transformation $\ga\mapsto\ga+\dee{\gl}$ is $\vph\mapsto\vph-\bdry_{t}\gl$,
$\v A\mapsto\v A+\nabla\gl$. The identity $\dee{\mathfrak{f}}=0$
takes the form of the homogeneous Maxwell equations, 
\begin{equation}
\nabla\cdot\v B=0,\qquad\nabla\cross\v E+\bdry_{t}\v B=0.\label{eq:homMaxwell}
\end{equation}
Indeed, by (\ref{eq:R3idf}), (\ref{eq:R3idg}) and $\dee t\wedge\dee t=0$,
\[
\dee{\mathfrak{f}}=\big((\nabla\cross\v E)\mathbin{\lrcorner}\dV\big)\wedge\dee t+(\nabla\cdot\v B)\dV+\dee t\wedge(\bdry_{t}\v B\mathbin{\lrcorner}\dV),
\]
and the last term equals $(\bdry_{t}\v B\mathbin{\lrcorner}\dV)\wedge\dee t$.

We write the free current, a 3-form, and the dipole form, a 2-form,
as 
\begin{equation}
\mathfrak{J}_{f}=\rho_{f}\dV-(\v j_{f}\mathbin{\lrcorner}\dV)\wedge\dee t,\qquad\mathfrak{g}=\v P\mathbin{\lrcorner}\dV+\v M^{\flat}\wedge\dee t,\label{eq:Jg4}
\end{equation}
with time-dependent fields $\rho_{f}$, $\v j_{f}$, $\v P$ and $\v M$
on $\rthree$. Every 3-form and every 2-form on $\mathcal{X}$ can
be written in this way, in a unique way. A direct computation, using
(\ref{eq:R3id})\textendash (\ref{eq:R3idb}) and $\dee t\wedge\dee t=0$,
gives (see Appendix~\ref{app:vec}), for every function $\rho$ and
vector fields $\v j$, $\v a$ and $\v b$, 
\begin{align}
\big(\rho\dV-(\v j\mathbin{\lrcorner}\dV)\wedge\dee t\big)\wedge\big(-\vph\dee t+\v A^{\flat}\big) & =\big(\rho\vph-\v j\cdot\v A\big)\dee t\wedge\dV,\label{eq:id1}\\
\big(\v a\mathbin{\lrcorner}\dV+\v b^{\flat}\wedge\dee t\big)\wedge\big(\v E^{\flat}\wedge\dee t+\v B\mathbin{\lrcorner}\dV\big) & =-\big(\v a\cdot\v E+\v b\cdot\v B\big)\dee t\wedge\dV.\label{eq:id2}
\end{align}
For example, in (\ref{eq:id1}), 
\[
\rho\dV\wedge(-\vph\dee t)=-\rho\vph\,\dV\wedge\dee t=\rho\vph\,\dee t\wedge\dV,
\]
since moving $\dee t$ across the 3-form $\dV$ gives the sign $(-1)^{3}$,
and 
\[
-(\v j\mathbin{\lrcorner}\dV)\wedge\dee t\wedge\v A^{\flat}=(\v j\mathbin{\lrcorner}\dV)\wedge\v A^{\flat}\wedge\dee t=(\v j\cdot\v A)\dV\wedge\dee t=-(\v j\cdot\v A)\dee t\wedge\dV.
\]
The other terms vanish, because they contain either $\dee t$ twice
or a 4-form in the three variables $x^{i}$.

\subsection{Energy, free and bound currents}

By (\ref{eq:U}), (\ref{eq:id1}) and (\ref{eq:id2}), the energy
of a spacetime region $\mathcal{R}$ takes the form 
\begin{equation}
U_{\mathcal{R}}(\ga)=\int_{\mathcal{R}}\big(\rho_{f}\vph-\v j_{f}\cdot\v A-\v P\cdot\v E-\v M\cdot\v B\big)\dee t\dV.\label{eq:U4}
\end{equation}
It is minus the interaction term of the Lagrangian of a polarizable
and magnetizable medium carrying free charge and current (see, e.g.,
{[}Jac99{]}), integrated over $\mathcal{R}$. For a cylinder $\mathcal{R}=[t_{1},t_{2}]\cross\gO$
and time-independent fields and currents, it is $t_{2}-t_{1}$ times
the sum of the electrostatic energy $\int_{\gO}(\rho_{f}\vph-\v P\cdot\v E)\dV$
of {[}GMS26a{]} and the magnetostatic energy $-\int_{\gO}(\v j_{f}\cdot\v A+\v M\cdot\v B)\dV$
of {[}GMS26b{]}. By Theorem~\ref{thm:decomp}, every weakly gauge-invariant
variational stress has this form, with unique $\rho_{f}$, $\v j_{f}$,
$\v P$ and $\v M$. In components, a general variational stress has
$4+16$ components, and weak gauge invariance leaves the $4$ components
of $(\rho_{f},\v j_{f})$ and the $6$ components of $(\v P,\v M)$.

The bound and total currents (\ref{eq:Jb}) are computed from 
\begin{equation}
\dee{\mathfrak{g}}=(\nabla\cdot\v P)\dV+\big((\bdry_{t}\v P+\nabla\cross\v M)\mathbin{\lrcorner}\dV\big)\wedge\dee t.\label{eq:dg4}
\end{equation}
Indeed, $\dee{(\v P\mathbin{\lrcorner}\dV)}=(\nabla\cdot\v P)\dV+\dee t\wedge(\bdry_{t}\v P\mathbin{\lrcorner}\dV)$,
and $\dee t\wedge(\bdry_{t}\v P\mathbin{\lrcorner}\dV)=(\bdry_{t}\v P\mathbin{\lrcorner}\dV)\wedge\dee t$,
since $\dee t$ commutes with 2-forms. In addition,
\[
\dee{(\v M^{\flat}\wedge\dee t)}=\dee{\v M^{\flat}}\wedge\dee t=\big((\nabla\cross\v M)\mathbin{\lrcorner}\dV\big)\wedge\dee{t,}
\]
 where the $\dee t$ component of $\dee{\v M^{\flat}}$ drops out.
Hence,
\[
\mathfrak{J}_{b}=\rho_{b}\dV-(\v j_{b}\mathbin{\lrcorner}\dV)\wedge\dee t\qquad\text{and}\qquad\mathfrak{J}=\rho\dV-(\v j\mathbin{\lrcorner}\dV)\wedge\dee t,
\]
 with 
\begin{equation}
\rho_{b}=-\nabla\cdot\v P,\qquad\v j_{b}=\bdry_{t}\v P+\nabla\cross\v M,\qquad\rho=\rho_{f}+\rho_{b},\qquad\v j=\v j_{f}+\v j_{b}.\label{eq:bound4}
\end{equation}
These are the familiar bound charge and bound current, including the
polarization current $\bdry_{t}\v P$ {[}Jac99, Zan13{]}. In the same
way, 
\begin{equation}
\dee{\mathfrak{J}_{f}}=\big(\bdry_{t}\rho_{f}+\nabla\cdot\v j_{f}\big)\dee t\wedge\dV,\label{eq:dJ4}
\end{equation}
and similarly for $\mathfrak{J}_{b}$ and $\mathfrak{J}$. The bound
current is conserved identically,
\[
\bdry_{t}\rho_{b}+\nabla\cdot\v j_{b}=-\nabla\cdot\bdry_{t}\v P+\nabla\cdot\bdry_{t}\v P+\nabla\cdot(\nabla\cross\v M)=0.
\]

The traction stress $\ga\mapsto\mathfrak{g}\wedge\ga$ gives the surface
currents. On the lateral boundary $[t_{1},t_{2}]\cross\bdry\gO$ of
a cylinder, 
\begin{equation}
\int_{[t_{1},t_{2}]\cross\bdry\gO}\mathfrak{g}\wedge\ga=\int^{t_{2}}_{t_{1}}\int_{\bdry\gO}\big((\v P\cdot\v{\nor})\vph-(\v M\cross\v{\nor})\cdot\v A\big)\dA\dee t,
\end{equation}
the energy of the bound surface charge $\v P\cdot\v{\nor}$ and of
the bound surface current $\v M\cross\v{\nor}$. On the time slices,
$\mathfrak{g}\wedge\ga$ restricts to $(\v P\cdot\v A)\dV$. These
terms account for the polarization current $\bdry_{t}\v P$ in the
principle of virtual power (\ref{eq:pvpnew}).

\subsection{Gauge invariance and conservation of charge}

By (\ref{eq:gaugechange2}) and (\ref{eq:id1}), a gauge transformation
changes the energy by 
\begin{equation}
\int_{\mathcal{R}}\mathfrak{J}_{f}\wedge\dee{\gl}=-\int_{\mathcal{R}}\big(\rho_{f}\bdry_{t}\gl+\v j_{f}\cdot\nabla\gl\big)\dee t\dV,\label{eq:gauge4}
\end{equation}
where we used (\ref{eq:id1}) with $\vph$ replaced by $-\bdry_{t}\gl$
and $\v A$ replaced by $\nabla\gl$, since $\dee{\gl}=\bdry_{t}\gl\,\dee t+(\nabla\gl)^{\flat}$.
For $\gl=-ct$, i.e., $\vph\mapsto\vph+c$, and static charges in
a cylinder, this is $(t_{2}-t_{1})\,c\int_{\gO}\rho_{f}\dV$, the
electrostatic result.

The three levels of gauge invariance of Section~\ref{sec:levels}
are now the following. Weak gauge invariance is the representation
(\ref{eq:U4}) of the energy. By Remark~\ref{rem:charge} and (\ref{eq:dJ4}),
interior gauge invariance, i.e., invariance of the energy of every
region of spacetime under the gauge transformations that vanish near
its boundary, is equivalent to the conservation of free charge, 
\begin{equation}
\bdry_{t}\rho_{f}+\nabla\cdot\v j_{f}=0.\label{eq:charge4}
\end{equation}
This can also be seen directly from (\ref{eq:gauge4}). If $\gl$
vanishes in a neighborhood of $\bdry\mathcal{R}$, integration by
parts in $t$ and in $\v x$ gives $\int_{\mathcal{R}}\gl\,(\bdry_{t}\rho_{f}+\nabla\cdot\v j_{f})\dee t\dV$.
Since the bound charge is conserved identically, (\ref{eq:charge4})
is equivalent to the conservation of the total charge, $\bdry_{t}\rho+\nabla\cdot\v j=0$.
Finally, by Theorem~\ref{thm:main}, the energy is gauge invariant
on every region if and only if $\rho_{f}=0$ and $\v j_{f}=0$, and
then $U_{\mathcal{R}}(\ga)=-\int_{\mathcal{R}}(\v P\cdot\v E+\v M\cdot\v B)\dee t\dV$,
which depends on the potential only through the field.

By Remark~\ref{rem:Mf}, a conserved free current on $\mathcal{X}$
is of the form $\mathfrak{J}_{f}=-\dee{\mathfrak{g}_{f}}$. By (\ref{eq:dg4}),
with $\mathfrak{g}_{f}=\v P_{f}\mathbin{\lrcorner}\dV+\v M^{\flat}_{f}\wedge\dee t$,
this means 
\[
\rho_{f}=-\nabla\cdot\v P_{f},\qquad\v j_{f}=\bdry_{t}\v P_{f}+\nabla\cross\v M_{f}.
\]
Every conserved charge and current distribution is thus the bound
charge and current of some polarization and magnetization. These are
not determined by the free current, and they expose surface currents
on cuts that the free current does not have.

\subsection{Mechanical forces}

Let the currents be embedded in matter in the sense of Section~\ref{sec:mech}.
The free current and the dipole form are convected as forms with the
world tubes of the material points, and a virtual motion is a deformation
of the family of world tubes, with the virtual velocity/displacement
$w=w^{0}\bdry_{t}+\v w$, where $\v w$ is the spatial part. By (\ref{eq:faraday}),
(\ref{eq:R3idc}), (\ref{eq:R3idd}) and $\v w\mathbin{\lrcorner}\dee t=0$
(see Appendix~\ref{app:vec}), 
\begin{align}
w\mathbin{\lrcorner}\mathfrak{f} & =(\v E\cdot\v w)\dee t-w^{0}\v E^{\flat}-(\v w\cross\v B)^{\flat},\label{eq:wf4}\\
w\mathbin{\lrcorner}\ga & =\v A\cdot\v w-\vph\,w^{0}.\label{eq:wa4}
\end{align}
For a 3-form $\mathfrak{J}=\rho\dV-(\v j\mathbin{\lrcorner}\dV)\wedge\dee t$,
(\ref{eq:id1}), with $-\vph$ replaced by $\v E\cdot\v w$ and $\v A$
replaced by $-w^{0}\v E-\v w\cross\v B$, gives 
\begin{equation}
-\mathfrak{J}\wedge(w\mathbin{\lrcorner}\mathfrak{f})=\Big(\big(\rho\v E+\v j\cross\v B\big)\cdot\v w-(\v j\cdot\v E)\,w^{0}\Big)\dee t\wedge\dV,\label{eq:lorentz4}
\end{equation}
where we used $\v j\cdot(\v w\cross\v B)=-(\v j\cross\v B)\cdot\v w$.
By Proposition~\ref{prop:mech}, for a conserved free current, 
\begin{equation}
\begin{split}F_{\mathcal{R}}(w)= & \int_{\mathcal{R}}\Big(\big(\rho\v E+\v j\cross\v B\big)\cdot\v w-(\v j\cdot\v E)\,w^{0}\Big)\dee t\dV\\
 & -\int_{\bdry\mathcal{R}}\mathfrak{g}\wedge(w\mathbin{\lrcorner}\mathfrak{f})+\int_{\bdry\mathcal{R}}\mathfrak{J}_{f}\wedge(w\mathbin{\lrcorner}\ga),
\end{split}
\label{eq:F4}
\end{equation}
with the total charge and current (\ref{eq:bound4}).

The body force is the Lorentz force density on the total charge and
current, free and bound, the latter including the polarization and
magnetization currents. Its time component, the coefficient of $w^{0}$,
is $-\v j\cdot\v E$, minus the rate at which the field does work
on the charges. The pair $(\rho\v E+\v j\cross\v B,-\v j\cdot\v E)$
is the covariant Lorentz force density. Since $-\mathfrak{J}\wedge(w\mathbin{\lrcorner}\mathfrak{f})=(w\mathbin{\lrcorner}\mathfrak{f})\wedge\mathfrak{J}$,
the body force has exactly the form $(w\mathbin{\lrcorner}\mathfrak{f})\wedge\mathfrak{J}$
in which the Lorentz force density is postulated in premetric electrodynamics
(see {[}Seg23{]}, {[}HO03{]}). Here, it follows from the energy (\ref{eq:U4})
and the embedding of the currents, and it acts on the total current.

The surface terms of (\ref{eq:F4}) are the force on the surface currents
exposed by the dipoles on $\bdry\mathcal{R}$, and the gauge-dependent
flux of potential energy-momentum of Section~\ref{sec:forcesystem}.
For a cylinder $\mathcal{R}=[t_{1},t_{2}]\cross\gO$, the boundary
consists of the two time slices, with the orientations $\pm\dV$,
and of the lateral boundary, on which $(\v j\mathbin{\lrcorner}\dV)\wedge\dee t$
restricts to $-(\v j\cdot\v{\nor})\,\dA\,\dee t$ with respect to
the induced orientation. With (\ref{eq:wa4}), the last term of (\ref{eq:F4})
is 
\begin{equation}
F^{g}_{\mathcal{R}}(w)=\Big[\int_{\gO}\rho_{f}\,(\v A\cdot\v w-\vph\,w^{0})\dV\Big]^{t_{2}}_{t_{1}}+\int^{t_{2}}_{t_{1}}\int_{\bdry\gO}(\v j_{f}\cdot\v{\nor})(\v A\cdot\v w-\vph\,w^{0})\dA\dee t.\label{eq:Fg4}
\end{equation}
The first term is the increase, between $t_{1}$ and $t_{2}$, of
the potential energy-momentum in $\gO$, the covector $\rho_{f}\ga$,
with components $(-\rho_{f}\vph,\rho_{f}\v A)$, paired with $(w^{0},\v w)$,
and the second is the amount that flows out through $\bdry\gO$ with
the free current. In agreement with the discussion of Section~\ref{sec:forcesystem},
$F^{g}_{\mathcal{R}}$ is the net potential energy-momentum that leaves
the spacetime region $\mathcal{R}$ with the free current. Both terms
involve the potentials, and both change under gauge transformations,
but, by Proposition~\ref{prop:forcegauge}, their sum is gauge invariant
whenever the free current does not cross $\bdry\mathcal{R}$.

For time-independent fields and currents, a time-independent spatial
virtual velocity, $w^{0}=0$, and a static gauge, the contributions
of the two time slices to (\ref{eq:F4}) cancel, and (\ref{eq:F4})
is $t_{2}-t_{1}$ times the sum of the electrostatic force of {[}GMS26a{]}
and the magnetostatic force of {[}GMS26b{]}. In particular, (\ref{eq:Fg4})
reduces to $t_{2}-t_{1}$ times the flux of potential momentum $\int_{\bdry\gO}(\v j_{f}\cdot\v{\nor})(\v A\cdot\v w)\dA$
of magnetostatics.

\section{Concluding remarks}

\label{sec:concl}

\textbf{Summary.} Geometric stress theory of {[}Seg02, Seg23{]}, with
$p$-forms as generalized velocities/displacements, gives a uniform
account of electrostatics, magnetostatics and electromagnetism in
matter. A current system is a collection of linear functionals on
potentials, one for each region, and it is represented by a variational
stress. Gauge transformations play the role of rigid motions. Weak
gauge invariance, the assumption that a gauge transformation changes
the energy only through the values of the gauge form, is the analog
of the symmetry of the stress, and it holds if and only if the variational
stress is $\mathfrak{J}_{f}\wedge\ga+\eps\,\mathfrak{g}\wedge\dee{\ga}$.
This splits the currents uniquely into free and bound parts. The action
of the traction stress is $\mathfrak{g}\wedge\ga$, and the body force
is the total current. Interior gauge invariance is equivalent to the
conservation of the free current, which, in spacetime, is the conservation
of charge. Gauge invariance of the energy is the analog of the balance
of forces, and it characterizes the absence of free currents. For
currents embedded in matter, the force derived from the energy is
the Lorentz-type force on the total current, the force on the surface
current exposed by the dipoles, and a gauge-dependent term. The boundary
force is derived from a traction stress. By Cartan's formula, the
gauge-dependent term is a boundary term, which we interpret as the
flux of potential energy-momentum carried by the free current, so
that the mechanical force on the matter is the gauge-invariant part.
For $n=4$, $p=1$, this is standard electromagnetism in matter, including
the bound currents $\bdry_{t}\v P+\nabla\cross\v M$ and the time
component $-\v j\cdot\v E$ of the force density.

\appendix

\section{Coordinate representations}

\label{app:coord}

In this appendix, we write the objects of Sections~\ref{sec:prelim}\textendash \ref{sec:mech}
in a chart $(x^{1},\dots,x^{n})$ of $\mathcal{X}$. Where applicable,
we use the summation convention for repeated indices, except for multi-indices,
for which we write the sums explicitly. We use the notation 
\[
\dee x:=\dee{x^{1}}\wedge\cdots\wedge\dee{x^{n}},\qquad\dee{x_{i}}:=\bdry_{i}\mathbin{\lrcorner}\dee x=(-1)^{i-1}\dee{x^{1}}\wedge\cdots\wedge\widehat{\dee{x^{i}}}\wedge\cdots\wedge\dee{x^{n}},
\]
where the hat indicates an omitted factor, so that $\dee{x^{j}}\wedge\dee{x_{i}}=\gd^{j}_{i}\dee x$.

\textbf{Jets and variational stresses.} Let $\{e_{\ga}\}$, $\ga=1,\dots,m$,
be a local basis of sections of $W$, so that a section is written
as $w=w^{\ga}e_{\ga}$. A jet $A\in J^{1}W$ at $x$ has the coordinates
$(x^{i},a^{\ga},a^{\ga}_{i})$, and, for $A=j^{1}_{x}w$, $a^{\ga}=w^{\ga}(x)$
and $a^{\ga}_{i}=\bdry_{i}w^{\ga}(x)$. The jet projection is $\pi^{1}_{0}(A)=a^{\ga}e_{\ga}$,
and the vertical subbundle $V^{1}_{0}J^{1}W$ consists of the jets
with $a^{\ga}=0$. A variational stress $S$ has components $R_{\ga}$
and $S^{i}_{\ga}$, such that 
\begin{equation}
S\cdot A=\big(R_{\ga}a^{\ga}+S^{i}_{\ga}a^{\ga}_{i}\big)\dee x,\qquad S\cdot j^{1}w=\big(R_{\ga}w^{\ga}+S^{i}_{\ga}\bdry_{i}w^{\ga}\big)\dee x.\label{eq:Scoord}
\end{equation}
A traction stress $s$ has components $s^{i}_{\ga}$, such that $s\cdot w=s^{i}_{\ga}w^{\ga}\dee{x_{i}}$,
and a body force $b$ has components $b_{\ga}$, such that $b\cdot w=b_{\ga}w^{\ga}\dee x$.
Since $\dee{(s^{i}_{\ga}w^{\ga}\dee{x_{i}})}=\bdry_{j}(s^{i}_{\ga}w^{\ga})\dee{x^{j}}\wedge\dee{x_{i}}=\bdry_{i}(s^{i}_{\ga}w^{\ga})\dee x$,
\begin{equation}
\dee{(s\cdot w)}=\big(\bdry_{i}s^{i}_{\ga}\,w^{\ga}+s^{i}_{\ga}\,\bdry_{i}w^{\ga}\big)\dee x.\label{eq:dscoord}
\end{equation}
Hence, the variational stress $\mathfrak{d}(s)$ of (\ref{eq:dfrak})
has the components 
\[
R_{\ga}=\bdry_{i}s^{i}_{\ga},\qquad S^{i}_{\ga}=s^{i}_{\ga}.
\]
The traction stress of a variational stress has the components of
its first-order part, 
\begin{equation}
p_{s}(S)\cdot w=S^{i}_{\ga}w^{\ga}\dee{x_{i}}.\label{eq:pscoord}
\end{equation}
This is consistent with (P1) and (P2). A zeroth-order variational
stress has $S^{i}_{\ga}=0$, so that its traction stress vanishes,
and $p_{s}(\mathfrak{d}(s))=s$. By (\ref{eq:div}) and (\ref{eq:dscoord}),
\[
\mathrm{div}\,S\cdot w=\dee{(p_{s}(S)\cdot w)}-S\cdot j^{1}w=\big(\bdry_{i}S^{i}_{\ga}-R_{\ga}\big)w^{\ga}\dee x,
\]
so that the body force (\ref{eq:body}) has the components 
\begin{equation}
b_{\ga}=R_{\ga}-\bdry_{i}S^{i}_{\ga}.\label{eq:bcoord}
\end{equation}
For $\mathcal{X}=\rthree$ with Cartesian coordinates and $W=T\rthree$,
$S^{i}_{\ga}$ is the stress component $\gs_{\ga i}$, $R_{\ga}=r_{\ga}$,
and (\ref{eq:bcoord}) is $\v b=\v r-\nabla\cdot\v{\gs}$, as in {[}GMS26b,
Sect.~3{]}. On the boundary of a region, $\dee{x_{i}}$ restricts
to $\nor_{i}\dA$, and $s\cdot w$ restricts to $(\v{\gs}\v{\nor})\cdot\v w\dA$.
Under a change of chart, the derivatives $a^{\ga}_{i}$ transform
with terms that involve the values $a^{\ga}$, and the components
$R_{\ga}$ do not transform as the components of a section of $L(W,\Lambda^{n}T^{*}\mathcal{X})$.
This is why the decomposition of $S$ into a zeroth-order part and
a first-order part is not invariant {[}Seg23, Sect.~10.8{]}.

\textbf{Jets of forms.} For $W=\Lambda^{p}T^{*}\mathcal{X}$, we use
the local basis $\{\dee{x^{I}}\}$, with $I$ running over the increasing
multi-indices of length $p$. A jet has the coordinates $(x^{i},a_{I},a_{Ik})$,
and a variational stress has components $R^{I}$ and $S^{Ik}$, such
that 
\begin{equation}
S\cdot A=\Big(\sum_{I}R^{I}a_{I}+\sum_{I}\sum_{k}S^{Ik}a_{Ik}\Big)\dee x.\label{eq:Sformcoord}
\end{equation}
For an increasing multi-index $K$ of length $p+1$, an index $k\in K$,
and $I:=K\setminus\{k\}$, let $\mathrm{sgn}(k,I)=\pm1$ be defined
by $\dee{x^{k}}\wedge\dee{x^{I}}=\mathrm{sgn}(k,I)\,\dee{x^{K}}$,
i.e., it is the sign of the permutation that brings $(k,I)$ to increasing
order. By (\ref{eq:dtlocal}), the components of $\tilde{d}(A)$ are
\[
\tilde{d}(A)_{K}=\sum_{k\in K}\mathrm{sgn}(k,K\setminus\{k\})\,a_{(K\setminus\{k\})k},
\]
and a section $\hat{S}$ of $L(\Lambda^{p+1}T^{*}\mathcal{X},\Lambda^{n}T^{*}\mathcal{X})$
with components $\hat{S}^{K}$ acts in the form 
\[
\hat{S}\big(\tilde{d}(A)\big)=\sum_{K}\hat{S}^{K}\tilde{d}(A)_{K}\,\dee x=\sum_{K}\sum_{k\in K}\hat{S}^{K}\,\mathrm{sgn}(k,K\setminus\{k\})\,a_{(K\setminus\{k\})k}\,\dee x.
\]
Comparing the coefficients of $a_{Ik}$ with (\ref{eq:Sformcoord}),
we conclude that $S$ is of the form (\ref{eq:RShat}), i.e., it is
weakly gauge invariant, if and only if 
\begin{equation}
S^{Ik}=\begin{cases}
\mathrm{sgn}(k,I)\,\hat{S}^{\{k\}\cup I}, & k\notin I,\\
0, & k\in I,
\end{cases}\label{eq:weakcoord}
\end{equation}
for some components $\hat{S}^{K}$. In words, $S^{Ik}$ depends only
on the set $\{k\}\cup I$, up to the sign of the permutation. For
$n=3$ and $p=1$, $I=(i)$ is a single index, and (\ref{eq:weakcoord})
says that $S^{ik}$ is skew-symmetric, as in {[}GMS26b, Prop.~4.2{]}.

The forms $\mathfrak{J}_{f}=\sum_{L}(\mathfrak{J}_{f})_{L}\,\dee{x^{L}}$
and $\mathfrak{g}=\sum_{N}\mathfrak{g}_{N}\,\dee{x^{N}}$, with $|L|=n-p$
and $|N|=n-p-1$, of Theorem~\ref{thm:decomp}(iv) are related to
the components $R^{I}$ and $\hat{S}^{K}$ of the sections $R$ and
$\hat{S}$ of (\ref{eq:RShat}), since $R(a)=\mathfrak{J}_{f}\wedge a$
and $\hat{S}(b)=\eps\,\mathfrak{g}\wedge b$ for a $p$-form $a$
and a $(p+1)$-form $b$, by Lemma~\ref{lem:wedge}. Let $I^{c}$
denote the complement of a multi-index $I$ in $\{1,\dots,n\}$, written
in increasing order, and let $\mathrm{sgn}(I^{c},I)=\pm1$ be defined
by $\dee{x^{I^{c}}}\wedge\dee{x^{I}}=\mathrm{sgn}(I^{c},I)\dee x$.
Since $\dee{x^{L}}\wedge\dee{x^{I}}$ vanishes unless $L=I^{c}$,
\[
R^{I}=\mathrm{sgn}(I^{c},I)\,(\mathfrak{J}_{f})_{I^{c}},\qquad\hat{S}^{K}=\eps\,\mathrm{sgn}(K^{c},K)\,\mathfrak{g}_{K^{c}}.
\]

\textbf{Interior products and Lie derivatives.} It is convenient to
write a $p$-form with antisymmetric components, $\ga=\frac{1}{p!}\ga_{i_{1}\cdots i_{p}}\dee{x^{i_{1}}}\wedge\cdots\wedge\dee{x^{i_{p}}}$,
summed over all indices. Then 
\begin{align}
(\dee{\ga})_{i_{0}\cdots i_{p}} & =\sum^{p}_{r=0}(-1)^{r}\bdry_{i_{r}}\ga_{i_{0}\cdots\widehat{i_{r}}\cdots i_{p}},\label{eq:dcoord}\\
(w\mathbin{\lrcorner}\ga)_{i_{2}\cdots i_{p}} & =w^{k}\ga_{ki_{2}\cdots i_{p}},\label{eq:ipcoord}\\
(\mathcal{L}_{w}\ga)_{i_{1}\cdots i_{p}} & =w^{k}\bdry_{k}\ga_{i_{1}\cdots i_{p}}+\sum^{p}_{r=1}\ga_{i_{1}\cdots i_{r-1}ki_{r+1}\cdots i_{p}}\bdry_{i_{r}}w^{k}.\label{eq:liecoord}
\end{align}
The last formula follows by differentiating the pullback. If $\dis_{\tau}(x)=x+\tau w(x)+o(\tau)$,
then $\bdry_{i}\dis^{j}_{\tau}=\gd^{j}_{i}+\tau\bdry_{i}w^{j}+o(\tau)$,
and 
\[
(\dis^{*}_{\tau}\ga)_{i_{1}\cdots i_{p}}(x)=\ga_{j_{1}\cdots j_{p}}(\dis_{\tau}(x))\,\bdry_{i_{1}}\dis^{j_{1}}_{\tau}\cdots\bdry_{i_{p}}\dis^{j_{p}}_{\tau}.
\]
Differentiating at $\tau=0$ by the product rule gives (\ref{eq:liecoord}).
For a 1-form, for example, (\ref{eq:dcoord})\textendash (\ref{eq:liecoord})
give $(w\mathbin{\lrcorner}\dee{\ga})_{i}=w^{k}(\bdry_{k}\ga_{i}-\bdry_{i}\ga_{k})$
and $\bdry_{i}(w\mathbin{\lrcorner}\ga)=\ga_{k}\bdry_{i}w^{k}+w^{k}\bdry_{i}\ga_{k}$,
whose sum is $w^{k}\bdry_{k}\ga_{i}+\ga_{k}\bdry_{i}w^{k}=(\mathcal{L}_{w}\ga)_{i}$,
in agreement with Cartan's formula.

\textbf{The mechanical variational stress.} By (\ref{eq:ipcoord}),
the components of $w\mathbin{\lrcorner}\mathfrak{f}$ at a point are
linear in the values $w^{k}$, and, by (\ref{eq:dcoord}) and (\ref{eq:ipcoord}),
those of $\dee{(w\mathbin{\lrcorner}\ga)}$ and $\dee{(w\mathbin{\lrcorner}\mathfrak{f})}$
are linear in $w^{k}$ and $\bdry_{j}w^{k}$. For instance, for $p=1$,
$\dee{(w\mathbin{\lrcorner}\ga)}=(\ga_{k}\bdry_{j}w^{k}+w^{k}\bdry_{j}\ga_{k})\dee{x^{j}}$.
Thus, (\ref{eq:Sm}) defines a variational stress $S_{M}$ on $J^{1}T\mathcal{X}$,
with components $R_{k}$, the coefficients of $w^{k}$, and $S^{j}_{k}$,
the coefficients of $\bdry_{j}w^{k}$, as in (\ref{eq:Scoord}). For
$\mathcal{X}=\rthree$, they are given explicitly in Section~\ref{sec:R3}.

\section{Forms and vector analysis}

\label{app:vec}

We derive the identities (\ref{eq:R3id})\textendash (\ref{eq:R3idg})
and the restriction formula of Section~\ref{sec:R3}. Let $(x^{1},x^{2},x^{3})$
be Cartesian coordinates in $\rthree$ and $\dV=\dee{x^{1}}\wedge\dee{x^{2}}\wedge\dee{x^{3}}$.
By the definition of the interior product, $\v a\mathbin{\lrcorner}\dV=\dV(\v a,\cdot,\cdot)$,
so that 
\begin{equation}
\v a\mathbin{\lrcorner}\dV=a_{1}\,\dee{x^{2}}\wedge\dee{x^{3}}+a_{2}\,\dee{x^{3}}\wedge\dee{x^{1}}+a_{3}\,\dee{x^{1}}\wedge\dee{x^{2}}=\half\eps_{ijk}a_{i}\,\dee{x^{j}}\wedge\dee{x^{k}}.\label{eq:adV}
\end{equation}

\begin{enumerate}
\item[(a)] Using (\ref{eq:adV}) and $\dee{x^{2}}\wedge\dee{x^{3}}\wedge\dee{x^{1}}=\dee{x^{3}}\wedge\dee{x^{1}}\wedge\dee{x^{2}}=\dV$,
while the products with a repeated factor vanish, $(\v a\mathbin{\lrcorner}\dV)\wedge\v b^{\flat}=(a_{1}b_{1}+a_{2}b_{2}+a_{3}b_{3})\dV$,
which is (\ref{eq:R3id}).
\item[(b)] We have
\[
\begin{split}\v a^{\flat}\wedge\v b^{\flat} & =a_{i}b_{j}\,\dee{x^{i}}\wedge\dee{x^{j}},\\
 & =(a_{2}b_{3}-a_{3}b_{2})\dee{x^{2}}\wedge\dee{x^{3}}+(a_{3}b_{1}-a_{1}b_{3})\dee{x^{3}}\wedge\dee{x^{1}}\\
 & \qquad\qquad\qquad\qquad+(a_{1}b_{2}-a_{2}b_{1})\dee{x^{1}}\wedge\dee{x^{2}},
\end{split}
\]
which, by (\ref{eq:adV}), is (\ref{eq:R3idb}).
\item[(c)] By (\ref{eq:adV}), 
\[
\begin{split}\v v\mathbin{\lrcorner}(\v u\mathbin{\lrcorner}\dV) & =\half\eps_{ijk}u_{i}(v_{j}\,\dee{x^{k}}-v_{k}\,\dee{x^{j}}),\\
 & =\eps_{ijk}u_{i}v_{j}\,\dee{x^{k}},\\
 & =(\v u\cross\v v)_{k}\,\dee{x^{k}},
\end{split}
\]
which is (\ref{eq:R3idc}).
\item[(d)] $\v v\mathbin{\lrcorner}\v a^{\flat}=a_{i}\,\dee{x^{i}}(\v v)=a_{i}v_{i}$,
which is (\ref{eq:R3idd}), and $\dee f=\bdry_{i}f\,\dee{x^{i}}$
is (\ref{eq:R3ide}).
\item[(e)] $\dee{\v a^{\flat}}=\bdry_{j}a_{i}\,\dee{x^{j}}\wedge\dee{x^{i}}$.
The computation in (b), with $a_{j}b_{i}$ replaced by $\bdry_{j}a_{i}$,
gives 
\[
\begin{split}\dee{\v a^{\flat}}= & (\bdry_{2}a_{3}-\bdry_{3}a_{2})\dee{x^{2}}\wedge\dee{x^{3}}+(\bdry_{3}a_{1}-\bdry_{1}a_{3})\dee{x^{3}}\wedge\dee{x^{1}}\\
 & +(\bdry_{1}a_{2}-\bdry_{2}a_{1})\dee{x^{1}}\wedge\dee{x^{2}},
\end{split}
\]
which, by (\ref{eq:adV}), is $(\nabla\cross\v a)\mathbin{\lrcorner}\dV$,
i.e., (\ref{eq:R3idf}).
\item[(f)] By (\ref{eq:adV}), 
\[
\begin{split}\dee{(\v a\mathbin{\lrcorner}\dV)} & =\bdry_{1}a_{1}\,\dee{x^{1}}\wedge\dee{x^{2}}\wedge\dee{x^{3}}+\bdry_{2}a_{2}\,\dee{x^{2}}\wedge\dee{x^{3}}\wedge\dee{x^{1}}\\
 & \qquad+\bdry_{3}a_{3}\,\dee{x^{3}}\wedge\dee{x^{1}}\wedge\dee{x^{2}}\\
 & =(\nabla\cdot\v a)\dV,
\end{split}
\]
which is (\ref{eq:R3idg}).
\item[(g)] Let $\v t_{1},\v t_{2}$ be orthonormal tangent vectors to $\bdry\mathcal{R}$
at a point, such that $(\v{\nor},\v t_{1},\v t_{2})$ is positively
oriented, so that $\v t_{1}\cross\v t_{2}=\v{\nor}$ and $\dA(\v t_{1},\v t_{2})=1$
for the induced orientation. Then $(\v a\mathbin{\lrcorner}\dV)(\v t_{1},\v t_{2})=\dV(\v a,\v t_{1},\v t_{2})=\v a\cdot(\v t_{1}\cross\v t_{2})=\v a\cdot\v{\nor}$.
Hence, $\v a\mathbin{\lrcorner}\dV$ restricts to $(\v a\cdot\v{\nor})\dA$.
\end{enumerate}
For the time-dependent fields of Section~\ref{sec:st}, the exterior
derivative on $\reals\times\rthree$ has, in addition, the terms with
$\dee t$. By (e) and (f), 
\[
\begin{split}\dee{\v a^{\flat}} & =\dee t\wedge(\bdry_{t}\v a)^{\flat}+(\nabla\cross\v a)\mathbin{\lrcorner}\dV,\\
\dee{(\v a\mathbin{\lrcorner}\dV)} & =(\nabla\cdot\v a)\dV+\dee t\wedge(\bdry_{t}\v a\mathbin{\lrcorner}\dV),
\end{split}
\]
and $\dee t\wedge\psi=(-1)^{k}\psi\wedge\dee t$ for a $k$-form $\psi$.

\section*{Acknowledgment}

The authors are grateful to Claude (Anthropic), an AI system, which
contributed substantially to the development of the arguments, to
the derivations, and to the preparation of this manuscript, and to
colleagues with whom earlier versions of the manuscript were discussed.


\begin{thebibliography}{GMS26a}
\bibitem[DO14]{DO14} L.~Dorfmann and R.~W.~Ogden. \emph{Nonlinear
Theory of Electroelastic and Magnetoelastic Interactions}. Springer,
2014.

\bibitem[EM90]{EM90} A. C. Eringen, G. A. Maugin. \emph{Electrodynamics
of Continua I: Foundations and Solid Media}. Springer, 1990.

\bibitem[Fed69]{Fed69} H.~Federer. \emph{Geometric Measure Theory}.
Springer, 1969.

\bibitem[Fra12]{Fra12} T.~Frankel. \emph{The Geometry of Physics:
An Introduction}. Cambridge University Press, 3rd ed., 2012.

\bibitem[GMS26a]{GMS26a} V.~Gol'dshtein, W.~H.~Müller, R.~Segev.
Electrostatics as a stress theory in $\rthree$: free and bound charge,
gauge invariance, and mechanical interactions. arXiv:2608.12900 {[}math-ph{]},
2026.

\bibitem[GMS26b]{GMS26b} V.~Gol'dshtein, W.~H.~Müller, R.~Segev.
Magnetostatics as a stress theory in $\rthree$: free and bound currents,
gauge invariance, and mechanical interactions. arXiv:2609.34766 {[}math-ph{]},
2026.

\bibitem[GPS02]{GPS02} H.~Goldstein, C.~Poole, J.~Safko. \emph{Classical
Mechanics}. Addison-Wesley, 3rd ed., 2002.

\bibitem[HO03]{HO03} F.~W.~Hehl, Yu.~N.~Obukhov. \emph{Foundations
of Classical Electrodynamics: Charge, Flux, and Metric}. Progress
in Mathematical Physics, Vol.~33. Birkhäuser, 2003.

\bibitem[Jac99]{Jac99} J.~D.~Jackson. \emph{Classical Electrodynamics}.
Wiley, 3rd ed., 1999.

\bibitem[Kov00]{Kov00} Y. Kovetz. \emph{Electromagnetic Theory}.
Oxford University Press, 2000.

\bibitem[Mau88]{Mau88} G. A. Maugin. \emph{Continuum Mechanics of
Electromagnetic Solids}. North-Holland, 1988.

\bibitem[Nol59]{Nol59} W.~Noll. The foundations of classical mechanics
in the light of recent advances in continuum mechanics. In \emph{The
Axiomatic Method, with Special Reference to Geometry and Physics}
(L.~Henkin, P.~Suppes, A.~Tarski, eds.), North-Holland, 1959, 266\textendash 281.

\bibitem[dR84]{dR84} G.~de Rham. \emph{Differentiable Manifolds:
Forms, Currents, Harmonic Forms}. Springer, 1984.

\bibitem[Seg02]{Seg02} R.~Segev. Metric-independent analysis of
the stress-energy tensor. \emph{J. Math. Phys.} 43 (2002), 3220\textendash 3231.

\bibitem[Seg16]{Seg16} R.~Segev. Continuum mechanics, stresses,
currents and electrodynamics. \emph{Phil. Trans. R. Soc. A} 374 (2016),
20150174.

\bibitem[Seg23]{Seg23} R.~Segev. \emph{Foundations of Geometric
Continuum Mechanics}. Birkhäuser, 2023.

\bibitem[TT60]{TT60} C.~Truesdell, R.~A.~Toupin. The classical
field theories. In \emph{Handbuch der Physik}, Vol.~III/1 (S.~Flügge,
ed.), Springer, 1960, 226\textendash 793.

\bibitem[Whi57]{Whi57} H.~Whitney. \emph{Geometric Integration Theory}.
Princeton University Press, 1957.

\bibitem[Zan13]{Zan13} A.~Zangwill. \emph{Modern Electrodynamics}.
Cambridge University Press, 2013.

\end{thebibliography}
\end{document}